\documentclass[aps,prd,twocolumn,showpacs,superscriptaddress,groupedaddress]{revtex4}  % for review and submission
\usepackage{graphicx,amssymb,amstext,amsmath}% Include figure files
\usepackage{bm}% bold math
\usepackage{subfigure}
\usepackage{dcolumn}   % needed for some tables
\newcommand\etal{\textsl{et al. }}

\usepackage[extension=pdf,final,pagebackref=true,backref=page,colorlinks=true,final=true,hyperindex=true,verbose=false,ps2pdf=true,citecolor=blue]{hyperref}

\begin{document}

\title{Arm locking for space-based laser interferometry gravitational wave observatories}

\author{Yinan Yu}
\email[]{yinan@phys.ufl.edu}
\author{Shawn Mitryk}
\author{Guido Mueller}
\affiliation{Department of Physics, University of Florida, Gainesville, Florida 32611, USA}

\date{\today}

\begin{abstract}
Laser frequency stabilization is a critical part of the interferometry measurement system of space-based gravitational wave observatories such as the Laser Interferometer Space Antenna (LISA). Arm locking as a proposed frequency stabilization technique, transfers the stability of the long arm lengths to the laser frequency. The arm locking sensor synthesizes an adequately filtered linear combination of the inter-spacecraft phase measurements to estimate the laser frequency noise, which can be used to control the laser frequency. At the University of Florida we developed the hardware-based University of Florida LISA Interferometer Simulator (UFLIS) to study and verify laser frequency noise reduction and suppression techniques under realistic LISA-like conditions. These conditions include the variable Doppler shifts between the spacecraft, LISA-like signal travel times, optical transponders, realistic laser frequency and timing noise. We review the different types of arm locking sensors and discuss their expected performance in LISA. The presented results are supported by results obtained during experimental studies of arm locking under relevant LISA-like conditions. We measured the noise suppression  as well as initial transients and frequency pulling in the presence of Doppler frequency errors. This work has demonstrated the validity and feasibility of arm locking in LISA.
\end{abstract}

\pacs{04.80.Nn, 95.55.Ym, 07.87.+v, 07.60.Ly, 42.60.Mi, 42.60.Lh}
\maketitle

\section{Introduction}

\begin{figure}[tb]
	\centering
		\includegraphics[width=0.45\textwidth]{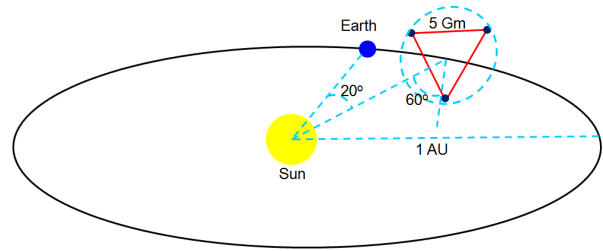}
	\caption{The heliocentric orbit of the LISA constellation. The constellation trails the Earth by $20^{\circ}$ and the plane of it is inclined with respect of the elliptic by $60^{\circ}$. The arm length between each two spacecraft is generally $5~\rm Gm$.}
	\label{fig:LISAOrbit}
\end{figure}

The Laser Interferometer Space Antenna (LISA) was a NASA/ESA collaborative space project and was supposed to be the first space-borne interferometric gravitational wave detector \cite{LISASRD}. LISA's goal was the detection of gravitational waves from astrophysical sources in the low frequency range between $3\times 10^{-5}~\rm Hz$ and $1~\rm Hz$. This frequency regime is rich in various gravitational wave sources including galactic binaries, massive black hole coalescences and extreme mass ratio inspirals (EMRIs). Following the demise of the collaboration, our European partners now plan a space-based LISA-like mission called eLISA or the Next Gravitational-wave Observatory (NGO) \cite{Amaro2012, NGO, YellowBookNGO}. eLISA is a candidate for ESA's L2 mission later in the next decade. NASA studies different mission concepts submitted to a Request for Information (RFI) and currently favors a mission which was submitted under the name SGO-mid \cite{SGO}, a three-arm/six-link version of the two-arm/4-link eLISA concept. NGO or eLISA and also SGO are LISA-like mission concepts with LISA-like frequency stabilization systems including the here discussed arm locking. Therefore we continue to use the LISA design as our reference design. This design consists of three spacecraft which form a near equilateral triangle with an average baseline of $5~\rm Gm$ (or $16.6~\rm s$ light travel time) as shown in Figure \ref{fig:LISAOrbit}. This constellation will be placed into a heliocentric orbit leading or trailing Earth by $20^{\circ}$. For us important is the fact that distortion of the gravitational potential caused by Earth and other planets cause relative motion between the spacecraft of up to $10~{\rm m/s}$ and changes in their distance on the order of $1\%$ of the nominal distances. 

Each spacecraft houses two drag-free proof masses that follow the geodesic motion and each proof mass is the end point of one LISA arm. A housing around the proof mass functions as a sensor to detect the relative position between the proof mass and the spacecraft. The Disturbance Reduction System (DRS) minimizes the acceleration of the poof mass due to undesired external forces and controls the thrusters on the spacecraft to track the geodesic motion. The Interferometric Measurement System (IMS) monitors changes in the separation between two proof masses on each respective spacecraft. Any modulation on the separation caused by gravitational waves and other residual spurious accelerations of the proof masses will be measured via interferometry with the desired sensitivity.

The interferometry uses a master/slave laser approach in which one of the lasers acts as the master laser and all other lasers will be offset phase-locked to this master laser. The offset frequency depends on the Doppler shifts of the laser fields; approximately $1~\rm MHz$ per $\rm m/s$ relative spacecraft motion for a $1~\rm {\mu m}$ laser, and is set by missions operations following a detailed frequency plan. Gravitational waves will then change the phases of the various laser beat signals taken on board of the three spacecraft. One of the key challenges of LISA interferometry is the reduction and cancellation of laser frequency noise in these beat signals. While the final cancellation will be achieved via time delay interferometry (TDI) \cite{Tinto1999, Armstrong1999, Shaddock2003, Mitryk2012}, it has been proposed to reduce the laser frequency noise by stabilizing the laser frequency to the LISA arms; the most stable references available in the LISA band \cite{Sheard2003}. Arm locking uses one or more beat signals formed on the master spacecraft to measure the frequency variations of the laser with respect to the LISA arms. In its original form, arm locking is comparable to stabilizing the frequency of a laser to an unequal arm Michelson interferometer. The difference to a standard Michelson interferometer based stabilization system is that the free spectral range of LISA's Michelson interferometer is in the $30~\rm mHz$ range ($1/16.6~\rm s$) and orders of magnitude smaller than the required bandwidth of the feedback loop. This imposes certain conditions on the shape of the loop which will be discussed in Section 2. In addition, the large time-dependent Doppler shift of the return beam adds a non-negligible contribution to the system; the end mirror in a standard Michelson interferometer in the typical optical lab is not moving with $10~\rm m/s$ towards or away from the beam splitter.

Initial proof of principle tests of arm locking \cite{Marin2005, Sheard2005} used much shorter sub-ms delays to demonstrate the basic idea. Our group developed a signal delay technique \cite{Thorpe2005a} and demonstrated arm locking using a few seconds delay. None of these experiments reached the $16.6~\rm s$ delay of LISA or added Doppler shifts to their experiment. Arm locking was further studied numerically and analytically by different groups \cite{McKenzie2009, Thorpe2011, Wand2009}. Their work included for example time varying Doppler shifts, the different clocks on the three spacecraft, and the spacecraft motion while we proceeded to set up the experiments to test arm locking under these realistic conditions. These experiments include tests of filtered linear combinations of the sensor signals from both arms which had been developed to increase the gain in the LISA band and to handle uncertainties and time variations of the Doppler shifts. In this paper, we report on these experimental results.

We will discuss several different arm locking schemes, discuss their advantages and shortcomings, and present several experimental results confirming our very good understanding of arm locking. In Section II we will first briefly review the architecture of LISA's long arm interferometry and its heterodyne phase measurements. Then we will give a realistic and generic arm locking model taking into account the optical transponders, realistic noise sources and laser frequency changes caused by the Doppler shifts. We will also introduce and characterize various arm locking sensor designs, including the single, common, dual and modified dual arm locking sensors. We will analyze their properties and limitations. Section III will deliberately describe the essential experimental components that constitutes various arm locking demonstrations in our experimental tests. The measurement results in Section IV will quantitatively verify the performance of single, common, dual and modified dual arm locking. We investigate the arm locking performance when it is combined with cavity pre-stabilization and when the noise sources in the optical transponders are not negligible. Section V will discuss more measurement details on the topics of initial transients in the lock acquisition process and Doppler frequency pullings in the steady state. The conclusion will be given in Section VI.

\section{Interferometry and Arm Locking}

The interferometry between the two proof masses forming one arm of the interferometer is split up into three sub-interferometer. Two local interferometer measure the changes of the positions of the two proof masses with respect to the optical bench on their host spacecraft while the third interferometer measures the changes in the distance between the two optical benches on the two spacecraft. Each of these interferometers is a heterodyne interferometer where the phase evolution of a laser beat signal is tracked and compared to other laser beat signals. A linear combination of these three measurements can then be used to calculate changes in the distance between the two proof masses. The spacecraft motion is tied to its proof mass motion using capacitive sensors and micro-Newton thrusters which steer the spacecraft around the free falling proof mass; this concept will be tested in ESA's upcoming LISA pathfinder mission \cite{Antonucci2011}. The residual motion between spacecraft and proof mass is expected to be in the few $\rm nm/\sqrt{Hz}$ range while the proof mass to proof mass motion is expected to be in the few $\rm pm/\sqrt{Hz}$ range increasing with $f^{-2}$ at frequencies below about $3~\rm mHz$ \cite{YellowBook}. The second arm of the interferometer will be measured the same way while the two lasers on board of each spacecraft will be phase-locked to each other. This phase lock creates an artificial beam splitter and allows to cancel in post-processing the laser frequency noise from all signals using TDI \cite{Mitryk2012, YellowBook}. Arm locking synthesizes the error signal from the inter-spacecraft phase measurements and its performance will be limited by the residual relative spacecraft motion.

The capability of TDI to cancel laser frequency noise is limited by uncertainties in the light travel time between the spacecraft. Minimizing this uncertainty is the task of the ranging system. For LISA, it was assumed that it is possible to measure the light travel time with an uncertainty of $\sim 3~\rm ns$ or a ranging error or $L \sim 1~\rm m$ \cite{YellowBook, Mitryk2012}. With this ranging accuracy and an allocated equivalent displacement noise of around $1~\rm pm/\sqrt{Hz}$, the required laser frequency noise for LISA and LISA-like missions is around
\begin{equation}
\delta \nu_{\rm pre-TDI} (f) < 300~\rm Hz~Hz^{-1/2} \sqrt{1 + \left(3~\rm mHz/f \right)^4} 
\end{equation}
pending on the noise allocation and specific mission design. To meet this requirement, the laser frequency has to be stabilized to a reference. This could be an optical cavity \cite{Drever1983, Mueller2005}, an unequal-arm Mach-Zehnder interferometer \cite{Heinzel2005} or a molecular line \cite{whitepaper}. In comparison, the LISA long arm is another reference which provides a quantitatively better stability in LISA's frequency band \cite{Sheard2003}. Arm locking allows to use already existing sensing signals and the control system can be fully implemented in on-board data processing units and no additional hardware is needed.
%LISA consists of three spacecraft in independent heliocentric orbits, arranged in a quasi-static equilateral triangular constellation, as shown in Figure \ref{fig:LISAOrbit} \cite{Jennrich2009}. The arm length between each two spacecraft is approximately $5\times 10^9~\rm m$, whereas due to the independence of each spacecraft orbit and the gravity field of the planets (most notably Earth), the arm length varies by up to $\pm 1\%$ over a year. The relative velocity between spacecraft Doppler shifts the frequency of the received laser used for interferometry by up to $20~\rm MHz$. 

%Each spacecraft houses two drag-free proof masses that follow the geodesic motion and each proof mass is the end point of one LISA arm. A housing around the proof mass functions as a sensor to detect the relative position between the proof mass and the spacecraft. The Disturbance Reduction System (DRS) minimizes the acceleration of the proof mass due to undesired external forces and controls the thrusters on the spacecraft to track the geodesic motion. The Interferometric Measurement System (IMS) monitors changes in the separation between two proof masses on each respective spacecraft. Any modulation on the separation caused by gravitational waves and other residual spurious accelerations of the proof masses will be measured via interferometry with the desired sensitivity.
\subsection{Architecture}

\begin{figure}[tb]
	\centering
		\includegraphics[width=0.45\textwidth]{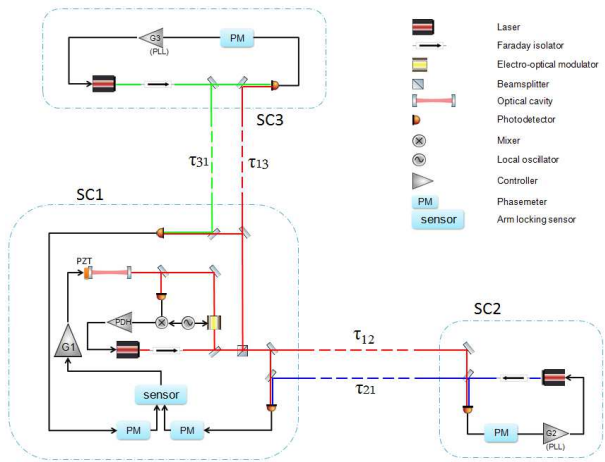}
	\caption{The baseline design of arm locking. The phasemeter on the far spacecraft $SC_i$ ($i = 2, 3$) measures the phase noise of the beat signal between the far laser $L_i$ and the incoming laser $L_1$. The measured phase difference is then used to phase lock the frequency of the laser $L_i$, as required by the optical transponders. The frequency noise of the $L_i$ is transmitted back to the master spacecraft $SC_1$ and then superimposed onto the instantaneous frequency noise of $L_1$. The phasemeters on the master spacecraft measure the two beat signals individually. Arm locking linearly combines these two measurements to control the frequency of the laser $L_1$. In this configuration the arm locking system is integrated with the laser $L_1$ pre-stabilized in a Pound-Drever-Hall setup with a length-tunable cavity.}
	\label{fig:LISAConstellation}
\end{figure}

Figure \ref{fig:LISAConstellation} shows the basic design of arm locking in LISA.  Laser $L_1$ on Spacecraft SC1 acts as the master laser to which the two lasers $L_i,\, i = 2, 3$, on $SC_i$ are phase-locked with an adjustable offset frequency in the $2 - 20~\rm MHz$ range. The photodetectors $PD_{i1}$ on SC1 measure the laser beat signals between $L_i$ and $L_1$. These beat signals are frequency shifted by the frequency offsets and Doppler shifts caused by the relative spacecraft motions. The signals are compared to reference signals which oscillate at the expected frequency of the beat signal using phasemeter:
\begin{equation}\label{Eq:PM_phase}
PM_{1i}(t) = \phi_1(t - \tau_i) - \phi_1(t) + \phi_i(t) + \phi_{N_i}(t) + \Delta \nu_{D_i}(t)t,
\end{equation}
where $\phi_1$ is the phase of $L_1$, $\phi_i(t)$ is the residual phase noise of $L_i$ after the phase locking servo is engaged and $\phi_{N_i}$ includes all additional noise sources such as clock noise and the residual spacecraft motion; in the following we will refer to this as sensor noise. $\Delta \nu_{D_i}$ is the Doppler error, a frequency offset that arises because of uncertainties in the Doppler shift. The initial single arm locking concept used one of these two signals as an error signal to stabilize the laser frequency to one of the LISA arms \cite{Sheard2003}. Two things deserve being pointed out: The difference between the two phases causes nulls in the transfer function (see Figure \ref{fig:SingleAL_Sensor}) of each individual arm at all Fourier frequencies equal to multiples of the free spectral range (FSR) $1/\tau_i$. At these nulls the phase in the transfer function changes from $-90^{\circ}$ to $+90^{\circ}$. This is well known from frequency stabilization systems based on unequal arm interferometer, but this has never been an issue as the bandwidth is usually kept well below the FSR to avoid unity gain oscillations. Sheard \etal  pointed out that a carefully designed controller with a gain roll off of less than $1/f$ maintains phase margin at these nulls and allows a much larger control bandwidth. However, the gain was still limited by the shallow roll off. By exploiting the pathlength mismatch between two long arms, dual arm locking uses both sensor signals to push the first sensor null to above the LISA frequency band and therefore achieves an almost flat response at low frequencies \cite{Sutton2008}. This flat response allows to increase the gain at lower frequencies much faster and significantly improve the performance of arm locking.

The second issue, the Doppler error was originally reported by Wand \etal  in meetings of the LISA Interferometry working group and later discussed in \cite{Wand2009, McKenzie2009, Thorpe2011}: As the last term in Equation 2 grows with time, the laser frequency will constantly change to track the accumulating phase change. This ramping of the laser frequency has to be reduced to ensure that all lasers stay within the single mode region of operation.

\subsection{Frequency Domain Analysis}

The performance of arm locking is best analyzed in the frequency domain. Taken the Laplace transform of Eq. \ref{Eq:PM_phase}, multiplying everything with $s$ to change from phase to frequency fluctuations ($\nu = s\phi$) and explicitly including the gain of the phase-locked loops at $SC_i$, the frequency noise relative to the reference signals is given by:
\begin{equation}
\nu_{1i}(s) = v_{L1}^0(s)P_{1i}(s) - \frac{1}{1+G_i}\nu_{L_i}^0(s) + \delta\nu_{N_i}(s) + \Delta\nu_{D_i}(s),
\end{equation}
where $\nu_{L1}^0(s)$ is the laser frequency noise of $L_1$ which arm locking will reduce. $\nu_{L_i}^0(s)$ is the free running laser frequency noise of the phase locked lasers. This noise is suppressed by the loop gain $G_i(s)$ of the phase lock loop. $\delta\nu_{N_i}(s)$ is the sensor noise and $\Delta\nu_{D_i}$ is the Doppler error.

\begin{equation}
P_{1i}(s) := 1 - \frac{G_i}{1 + G_i}e^{-s\tau_i}
\end{equation}
is the transfer function of the laser frequency noise of $L_1$ into the error signal. Note that for simplicity we have taken the liberty to write the noise terms as a linear sum. In reality, these noise terms are uncorrelated and have to be added quadratically.

Arm locking linearly combines the measured beat signals $\nu_{12}(s)$ and $\nu_{13}(s)$ to generate an error signal to control $L_1$. Following the notation used in Ref\cite{Sutton2008, McKenzie2009}, the arm locking sensor is
\begin{equation}\label{Eq:ALSensor}
H(s) = \boldsymbol{S_k}(s) \cdot \begin{bmatrix}P_{12}(s)\\P_{13}(s)\end{bmatrix},
\end{equation}
where $\boldsymbol{S_k}(s) = [h_+(s) + h_-(s), h_+(s) - h_-(s)]$ is known as the mapping vector. $h_+(s)$ and $h_-(s)$ are two filters placed in the common arm channel $P_+ = P_{12} + P_{13}$ and differential arm channel $P_- = P_{12} - P_{13}$ respectively. Their specific transfer functions depend on the desired arm locking configuration, which can be optimized based on the arm lengths, laser frequency noise, etc.

In steady state, the in-loop frequency noise of the arm locked laser $L_1$ is then given by
\begin{align}\label{Eq:RALM_LFN}
\nu_{L_1}(s) & = \nu_{L_1}^0(s) - \frac{G_1}{1 + G_1 H}[h_+ + h_-, h_+ - h_-] \cdot \begin{bmatrix}\nu_{12}(s)\\\nu_{13}(s)\end{bmatrix}\notag\\
						 & = \frac{1}{1 + G_1 H}\nu_{L_1}^0(s) + \frac{h_+ + h_-}{H}\delta \nu_{N}(s) \notag\\
						 & + \frac{h_+}{H}\Delta \nu_{\rm D+} + \frac{h_-}{H}\Delta \nu_{\rm D-}.
\end{align}

The first term of Eq. \ref{Eq:RALM_LFN} indicates the open-loop frequency noise suppressed by the open-loop gain $G_1 H$. The second term represents the noise sources which will limit arm locking performance. The last two terms represent the frequency pulling of the stabilized laser due to the common and differential Doppler frequency errors, which are defined as $\Delta \nu_{\rm D+} = \Delta \nu_{\rm D_2} + \Delta \nu_{\rm D_3}$ and $\Delta \nu_{\rm D-} = \Delta \nu_{\rm D_2} - \Delta \nu_{\rm D_3}$ respectively. In this notation, the Doppler frequency error enters the same way the residual spacecraft motion enters into the signal. In the LISA band, these two terms are several orders of magnitude smaller than the residual spacecraft motion and can be ignored for the in-band performance. However, these terms become dominant at frequencies around the orbital frequency ($\sim 1/\rm yr$).

\subsection{Sensor Characterization and Frequency Pulling}

Over the last years several different arm locking sensors characterized by their mapping vectors have been analyzed. In the following sections, we briefly review the main sensor configurations and discuss their expected in-band performance and the frequency pulling due to the Doppler error.

\subsubsection{Single arm locking}

\begin{figure}[tb]
	\centering
		\includegraphics[width=0.45\textwidth]{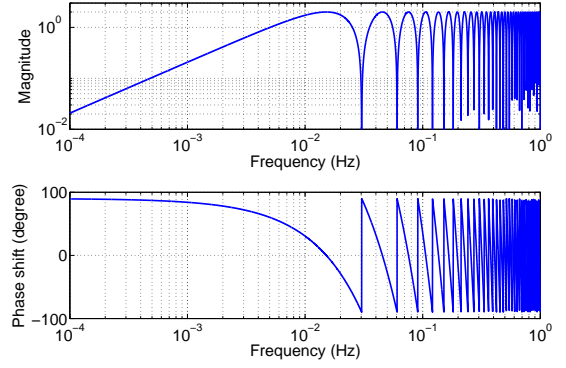}
	\caption{The magnitude and phase response of the single arm locking sensor ($\tau_2 = 33~\rm s$).}
	\label{fig:SingleAL_Sensor}
\end{figure}

The mapping vector for single arm locking is $\boldsymbol{S_k} = [1, 0]$; the interferometer output of only one arm is directly used as the sensor signal. The sensor transfer function is shown in Figure \ref{fig:SingleAL_Sensor} and simply given by:
%\begin{align}
%\nu_{L_1}(s) & = \nu_{L_1}^0(s) - \frac{G_1}{1 + G_1 H_{\rm S}}\nu_{12}(s) \notag\\
						 %& = \frac{1}{1 + G_1 H_{\rm S}}\nu_{L_1}^0(s) \notag\\
						 %& + \frac{G_1}{1 + G_1 H_{\rm S}} \delta \nu_{N_2}^{\prime}(s) + \frac{G_1}{1 + G_1 H_{\rm S}}\Delta \nu_{\rm D_2},
%\end{align}
%where the sensor transfer function $H_{\rm S}(s)$ is simply given by
\begin{equation}
H_{\rm S}(s) = P_{12}(s) = 1 - \frac{G_2}{1 + G_2} e^{-s\tau_2} \approx 1 - e^{-s\tau_2}.
\end{equation}

The nulls and phase jumps at Fourier frequencies $n/\tau_2$ are caused by the insensitivity of the interferometric setup to laser frequency noise at these frequencies. The transfer function also have multiple unity gain frequencies, on both sides near each null. These unity gain frequencies place additional stability constraints on the controller design, requiring a slope less steep than $1/f$ to avoid excessive phase shifts at these unity gain frequencies. Nevertheless, the large phase shift at sensor nulls still causes unavoidable noise amplifications, corresponding to multiple peaks in the closed-loop transfer function at these frequencies which include frequencies within the LISA measurement band. The limited gain in small bands around these nulls also causes oscillations or start-up transients in the laser frequency. The amplitude and decay time of these oscillations depend on the controller gain \cite{Tinto2004}.

We are also interested in the system response to the Doppler error $\Delta \nu_{\rm D_2}$. When the Fourier frequency is very low ($f \ll 1/\tau_2$), the sensor transfer function approximates to $s\tau_2$. Presumably, at low frequencies the arm locking open-loop yields the high gain limit $G_1 H_{\rm S} \gg 1$. Therefore, the transfer function $\frac{G_1}{1 + G_1 H_{\rm S}}$ yields $1/H_{\rm S} \approx 1/s\tau_2$. This indicates that a single arm locking loop integrates the transponder noise and also accumulates the Doppler error, causing a frequency pulling with an instantaneous rate given by
\begin{equation}
\left(\frac{\rm{d}\nu_{\rm{L}}}{\rm{d}t}\right)_{\rm S} = \frac{\Delta \nu_{\rm D_2}}{\tau_2}. 
\end{equation}
%As the round-trip light travel time is long on a single arm ($\approx 33~\rm s$), the noise limitations and Doppler frequency pullings can be adequately controlled.
%
%Although the idea of single arm locking has been theoretically and experimentally validated by multiple research groups \cite{Marin2005, Sheard2005, Thorpe2005b, Livas2009, Wand2009}, the shortcomings of single arm locking make it ultimately inappropriate for LISA. It is very noticeable that the first null shows up right inside the LISA band, which in any case is not an ideal noise suppression performance due to the noise amplifications at the nulls. Moreover, since the first null ($1/\tau_2 \approx 30~\rm mHz$) is in the LISA band, the gain around that region must be high enough to provide a decent noise suppression. Considering the controller will have a slope less steep than $1/f$, an extremely large bandwidth would be needed. The third disadvantage of single arm locking is the start-up transients caused by the non-zero initial error signal, which creates the repeating 33 seconds noise with damped oscillations. The amplitude of the transients can be effectively reduced using a ramping design, which increases the controller gain to the desired value gradually. However, this method cannot be used to reduce the relaxation time of the decaying transients. More details about the start-up transients can be seen in Section VI.

\subsubsection{Common arm locking}

Herz pointed out that exploring the arm length difference allows to reconstruct the laser frequency noise and to actively reduce the start-up transients \cite{Herz2005}. This idea became the seed of new sensor designs which incorporate the second arm. The simplest version uses the mapping vector $\boldsymbol{S_k}(s) = [1, 1]$ which generates the following transfer function ($s = i\omega$):
\begin{equation}
H_{\rm C}(s) = 2 - e^{-s\tau_2} - e^{-s\tau_3} = 2(1 - e^{-i\omega\bar{\tau}}\cos\omega\Delta\tau).
\end{equation}

\begin{figure}[tb]
	\centering
		\includegraphics[width=0.45\textwidth]{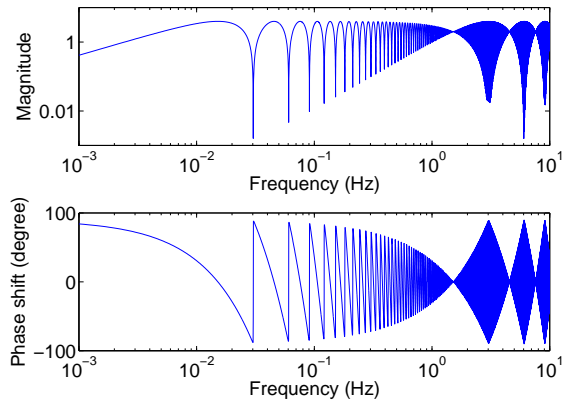}
	\caption{The magnitude and phase response of the common arm locking sensor in the case of $\sim 1\%$ arm length mismatch ($\bar{\tau} = 33~\rm s, \, \Delta \tau = 0.16~\rm s$).}
	\label{fig:CommonAL_Sensor}
\end{figure}

We refer to this sensor as common arm locking although the terminology is not completely consistent throughout the literature. This sensor transfer function with $\bar{\tau} \equiv (\tau_2 + \tau_3)/2 = 33~\rm s$ and $\Delta\tau \equiv (\tau_2 - \tau_3)/2 = 0.16~\rm s$ ($\sim 1\%$ arm length mismatch) is plotted in Figure \ref{fig:CommonAL_Sensor}. The magnitude of the transfer function does not decrease to zero when $f = n/\bar{\tau}$ unless $f$ is also an integer multiple of $1/\Delta \tau$. The figure also shows that the phase shift at $1/\bar{\tau} \approx 30~\rm mHz$ is still close to $90^{\circ}$. Similar to single arm locking, noise amplifications may also occur at the frequencies with large phase shift in common arm locking. This characteristic indicates that despite of the introduction of a second arm, common arm locking is essentially not a significant improvement of single arm locking in terms of the gain advantage and system stability.

At frequencies $f \ll 1/\bar{\tau}$, the sensor transfer function approximates to $2s\bar{\tau}$ and the Doppler frequency pulling rate is given by
\begin{equation}
\left(\frac{\rm{d}\nu_{\rm{L}}}{\rm{d}t}\right)_{\rm C} = \frac{\Delta \nu_{\rm D+}}{2\bar{\tau}}, 
\end{equation}
which is similar to single arm locking.

\subsubsection{Dual arm locking}

\begin{figure}[tb]
	\centering
		\includegraphics[width=0.45\textwidth]{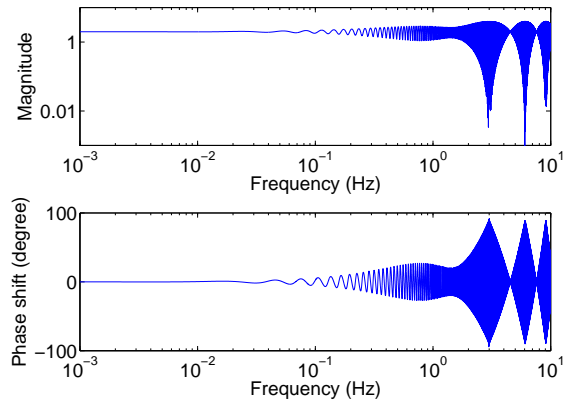}
	\caption{The magnitude and phase response of the dual arm locking sensor in the case of $\sim 1\%$ arm length mismatch ($\bar{\tau} = 33~\rm s, \, \Delta \tau = 0.16~\rm s$).}
	\label{fig:DualAL_Sensor}
\end{figure}

Sutton and Shaddock realized that emphasizing the arm length difference in the sensor signal would significantly reduce the rapid changes in the sensor response inside the LISA band \cite{Sutton2008}. Using the frequency dependent mapping vector $\boldsymbol{S_k}(s) = [1+\frac{E(s)}{s\Delta \tau}, 1-\frac{E(s)}{s\Delta \tau}]$ generates the following sensor signal:
\begin{equation}\label{Eq:DualSensor}
H_D(s = i\omega) = 2(1 - e^{-i\omega\bar{\tau}}\cos\omega\Delta\tau) + \frac{E(\omega)}{\omega\Delta\tau}2\sin\omega\Delta\tau.
\end{equation}
$E(s)$ is a low pass filter with a pole around $1/4\Delta\tau$. Through this combination, the delayed frequency information can be eliminated at low frequencies ($f \ll 1/\Delta \tau$) and the frequency response is therefore almost flat in the LISA band. In the high frequency region the integrated differential arm contributes excessive phase shift at sensor nulls (the first at $1/\Delta \tau$). This is compensated by a pole in E(s) which maintains the system stability by attenuating and phase-shifting the differential arm within the instability region. Without affecting the overall magnitude response, this filter ensures that the common arm dominates at all frequencies above $1/\Delta \tau$ such that the phase loss at sensor nulls is alleviated back to $90^{\circ}$. The magnitude and phase responses of the dual arm locking sensor are shown in Figure \ref{fig:DualAL_Sensor}, where the arm length mismatch is $\sim 1\%$ and the low-pass filter has a single pole at $1~\rm Hz$. The first impulse null of this dual arm locking sensor is at $1/\Delta \tau$, while at $n/2\Delta \tau$ the magnitude response has a local minimum.

Since $h_-(s) = \frac{E(s)}{s\Delta \tau}$ is an integrator and $H(s) \approx 2$ in the LISA band, the transponder noise and the differential Doppler frequency error will be accumulated in the differential channel. This becomes even more critical when the arm length mismatch is very small, which is the main disadvantage of dual arm locking. This is most obvious in the frequency pulling rate:
\begin{equation}
\left(\frac{\rm{d}\nu_{\rm{L}}}{\rm{d}t}\right)_{\rm Dual} = \frac{\Delta \nu_{\rm D-}}{2\Delta \tau},
\end{equation}
which now scales with $1/\Delta\tau \gg 1/\bar{\tau}$.

\subsubsection{Modified dual arm locking} 

Wand \etal and McKenzie \etal \cite{McKenzie2009} independently realized that the Doppler frequency pulling rate can be reduced if the common arm signal dominates again at frequencies below $1/\bar{\tau}$. This combination retains the overall flat transfer function of the dual arm locking sensor below $1/2\Delta \tau$ and effectively reduces the frequency pulling due to the differential Doppler frequency error. In general, the mapping vector of modified dual arm locking can be calculated using (see Eq. \ref{Eq:ALSensor}) 
\begin{align}
h_+(s) & = F_{\rm C}(s) + F_{\rm D}(s) \notag\\
h_-(s) & = \frac{E(s)F_{\rm D}(s)}{s\Delta \tau}
\end{align}
and the modified dual arm locking sensor is then given by
\begin{equation}
H_{\rm MD}(s) = F_{\rm C}(s)P_+(s) + F_{\rm D}(s)H_{\rm D}(s),
\end{equation}
where $H_{\rm D}(s)$ is the dual arm locking sensor signal given in Eq. \ref{Eq:DualSensor}.

Therefore, at frequencies below $1/\bar{\tau}$, $F_{\rm D}(s)$ needs to provide only limited gain while $F_{\rm C}(s)$ needs to provide high gain to amplify the common arm channel. Given such a design, the common arm channel will dominate in the low frequency range as well as above $1/2\Delta\tau$ while the differential arm signal dominates between $1/\bar{\tau}$ and $1/2\Delta\tau$.

\begin{figure}
	\centering
		\includegraphics[width=0.45\textwidth]{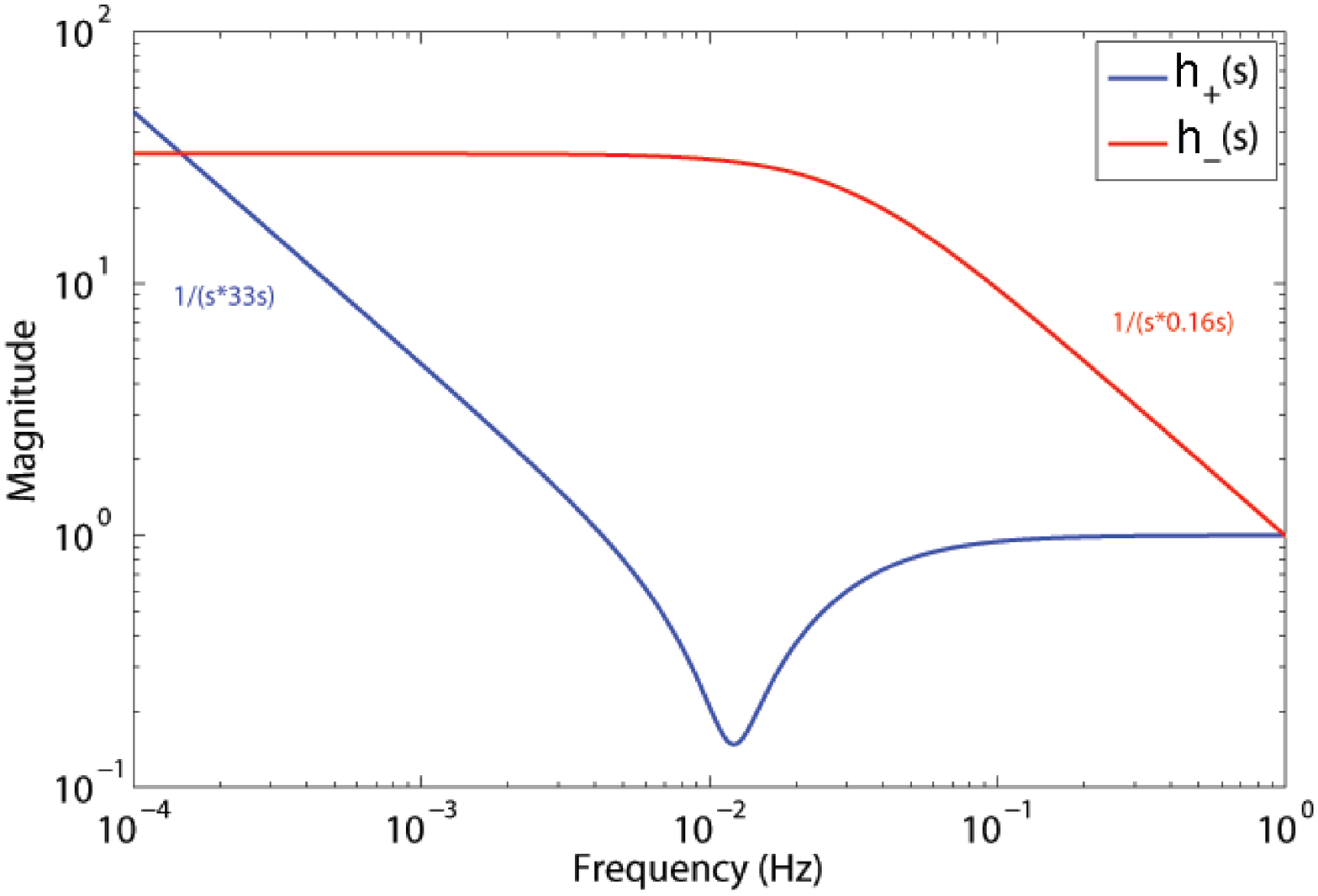}
	\caption{The magnitude response of $h_+(s)$ and $h_-(s)$ with $\bar{\tau} = 33~\rm s$ and $\Delta \tau = 0.16~\rm s$ for modified dual arm locking.}
	\label{fig:ModDualAL_HH}
\end{figure}

\begin{figure}[tb]
	\centering
		\includegraphics[width=0.45\textwidth]{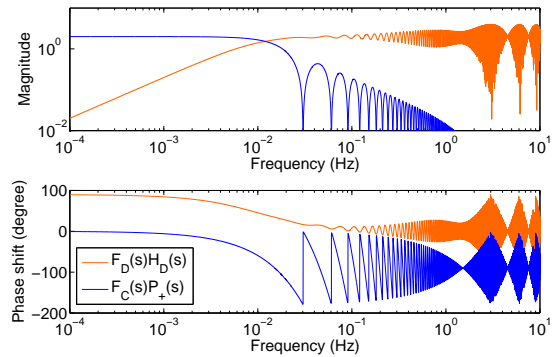}
	\caption{The magnitude and phase response of the dual and common arm components that constitute the modified dual arm locking sensor ($\sim 1\%$ arm length mismatch). The dual arm component is high-pass filtered and the common arm component is low-pass filtered. Therefore, the common arm component will dominate at the frequencies below $1/\bar{\tau}$ and the dual arm component will dominate at the frequencies above it. The phase shift of the common arm component in the low frequency region needs to be attenuate to zero to maintain the overall transfer function.}
	\label{fig:ModDualAL_Sensor}
\end{figure}

In this article we follow the design for the University of Florida LISA Interferometry Simulator (UFLIS) described in Ref \cite{Yu2011}, where the low-pass filter $F_{\rm C}(s)$ utilizes an integrator with a scaling factor of $1/\bar{\tau}$ and the high-pass filter $F_{\rm D}(s)$ has a zero at DC and a pole at about $1/\bar{\tau} \approx 30~\rm mHz$. This simplified design increases the efficiency of signal processing without significantly affecting the noise suppression performance of modified dual arm locking. Figure \ref{fig:ModDualAL_HH} plots the transfer functions of $h_+(s)$ and $h_-(s)$ with $\bar{\tau} = 33~\rm s$ and $\Delta \tau = 0.16~\rm s$ and Figure \ref{fig:ModDualAL_Sensor} shows the transfer functions of the dual and common arm components that constitute the modified dual arm locking sensor. 

The magnitude responses of $h_+(s)$ and $h_-(s)$ feature a slope of $1/s\bar{\tau}$ and $1/s\Delta \tau$, respectively. At low frequencies $f \ll 1/\bar{\tau}$, $h_-(s)$ flattens out and approaches a constant factor of $\bar{\tau}/(2\pi\Delta\tau)$. At high frequencies $f \gg 1/\bar{\tau}$, $h_+(s)$ flattens out and approaches $1$. Compared with the $1/s\Delta \tau$ slope that dominates in dual arm locking, the gain has been suppressed to either $\bar{\tau}/(2\pi\Delta\tau)$ or $1/s\bar{\tau}$, depending on which one is larger. Especially, the dominant frequency pulling rate, which is now completely attributed to the common Doppler error, is given by
\begin{equation}
\left(\frac{\rm{d}\nu_{\rm{L}}}{\rm{d}t}\right)_{\rm Mod\,Dual} = \frac{\Delta \nu_{\rm D+}}{2\bar{\tau}}.
\end{equation}

\subsection{External Noise Limitations}

In reality, the stabilized laser frequency noise is dominated by several external noise sources if we assume a standard DC-coupled arm locking controller. In addition to the technical noise (e.g., digitization noise) introduced by the arm locking system, a nominal noise source is the optical transponder noise coming from the constellation phase-locking loops. The optical transponder noise enters as sensing noise in the heterodyne phase detection (clock noise, shot noise, technical noise in phasemeters), as well as pathlength noise of the arm length reference (spacecraft motion).

\begin{itemize}
\item Clock noise - The phase of a beat signal is measured by comparing it to a timing reference (the local ultra-stable oscillator). Therefore, the acquired phase value is always relative to the phase noise of the referencing clock. The clock noise is proportional to the nominal frequency $\Omega$ of the measured beat signal, i.e.,
	\begin{equation}
	\delta \nu_{\rm clock}(f) = \Omega \cdot \delta \nu_{\rm clock}^{\rm Norm}(f),
	\end{equation}	
where	$\delta \nu_{\rm clock}^{\rm Norm}(f)$ are the fractional frequency fluctuations, corresponding to the normalized clock frequency noise at $1~\rm Hz$ clock frequency. The fractional frequency fluctuation is estimated to be approximately $2.4 \times 10^{-12}/\sqrt{f}~{\rm Hz^{-1/2}}$. The clock noises from phasemeters on the same spacecraft are correlated, while they are uncorrelated on different spacecraft.

\item Spacecraft motion - The LISA arm length is an excellent reference to stabilize the laser frequency. The stability of this length reference is limited by the residual spacecraft motion which is dragged by the DRS to track the proof mass motion. However, due to gain limitations in the DRS, the spacecraft cannot follow the proof mass perfectly and the residual motion is approximately
	\begin{equation}
	\delta L_{\rm SC} (f) = 1.5 \times 10^{-9} \sqrt{1 + \left( \frac{8~{\rm mHz}}{f} \right)^4}~{\rm m~Hz^{-1/2}}. 
	\end{equation}
	
The length uncertainty of one arm includes the spacecraft motions of two spacecraft at each end. This limited stability in the length reference will cause a phase noise given by $\delta \varphi_{\rm SC}(f) = \delta L_{\rm SC} (f)/\lambda_{\rm L}$ in the phasemeter measurements, where $\lambda_{\rm L}$ is the laser wavelength.
 
\item Shot noise - Limited number $N$ of photons per second received by the photodetectors. With $100~\rm pW$ light received at the photodetector the shot noise is given by
	\begin{align}
	\delta \varphi_{\rm shot} & = \frac{1}{2\pi\sqrt{N}} = \sqrt{\frac{\hbar c}{2\pi}\frac{1}{\lambda P}} \notag\\
														& = 6.9 \times 10^{-6}~{\rm cycles~Hz^{-1/2}}.
	\end{align}

The shot noise will be added to the signal at all photodetectors as a frequency noise with a phase-to-frequency conversion $\delta \nu_{\rm shot}(f) = 2\pi f \delta \varphi_{\rm shot}$.
	
\item Technical noise - Includes the ADC noise in the A/D conversion of the beat signal, as well as the finite precision of integer arithmetic, known as the digitization noise in the phasemeters. A digital signal with a sampling frequency of $f_s$ and a precision of N-bit generally carries the digitization noise (See also Section III.A.1 and Ref \cite{ThorpeT}.)

	\begin{equation}\label{Eq:Ndig}
	\delta \nu_{\rm dig} = \frac{f_{\rm clock} \cdot 2^{-N}}{\sqrt{6\cdot f_s}}.
	\end{equation}

The digitization noise is white and independent of frequency. Arm locking requires phasemeters with a fast data rate ($\sim 100~\rm kHz$) to maintain the bandwidth. One of the commonly used arm locking phasemeters in UFLIS, the 48-bit phasemeter, has a $62.5~\rm MHz$ clock frequency and a $488~\rm kHz$ data rate. With this design, the arm locking sensor will sense the digitization noise given by $1.3 \times 10^{-10}~{\rm Hz~Hz^{-1/2}}$.

\end{itemize}

\begin{figure}
	\centering
		\includegraphics[width=0.45\textwidth]{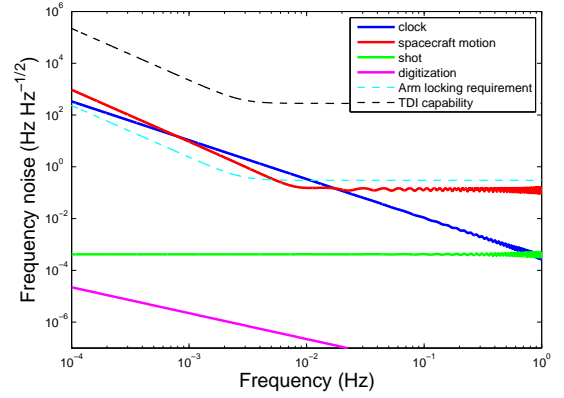}
	\caption{\label{fig:DAL_noise} The noise floors in dual arm locking are sensitive to the arm length mismatch. Here we assume a relatively short arm length mismatch of $0.1\%$ and the noise floors are significantly higher than initially recommended for LISA using arm locking. As the noise floor in dual arm locking is inversely proportional to the arm length mismatch, the performance of dual arm locking is insufficient to meet the TDI capability when the arm length mismatch is less than about $60~\rm km$.}
\end{figure}

\begin{figure}
	\centering
		\includegraphics[width=0.45\textwidth]{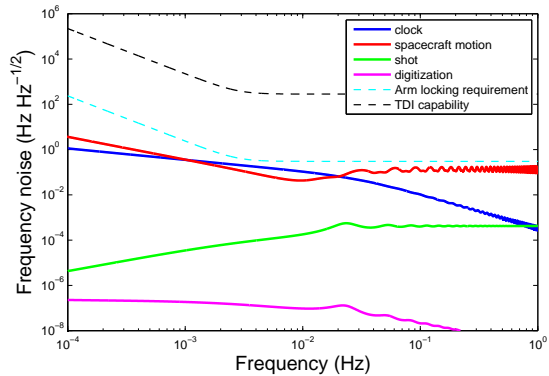}
	\caption{\label{fig:MDAL_noise} The noise floors in modified dual arm locking are less dependent on the arm length mismatch due to the low-frequency filtering scheme. The noise floors are effectively suppressed by the high-pass filter at frequencies below $1/\bar{\tau}$. With a more carefully designed high-pass filter, the noise floors will be further reduced and asymptotically approach the noise floors determined in the common arm locking configuration.}
\end{figure}

The impacts of these noise sources on the arm locking performance can be evaluated using Eq. \ref{Eq:RALM_LFN}, where the second term represents the noise contributions to the frequency noise of the output laser. In the cases of the dual and modified dual arm locking sensors described above, the expected arm locking performances are plotted in Figure \ref{fig:DAL_noise} and Figure \ref{fig:MDAL_noise}. Here the arm length mismatch is assumed to be as short as $0.1\%$, corresponding to a $0.016~\rm s$ differential time delay \cite{McKenzie2009}. For the dual arm locking configuration, Eq. \ref{Eq:RALM_LFN} indicates that the noise floor is primarily composed of the integration of the quadrature sum of the uncorrelated noises on both arms and inversely proportional to the arm length mismatch. As shown in Figure \ref{fig:DAL_noise}, the clock noise and spacecraft motion as the dominant noise floor prevent arm locking from meeting the recommended arm locking objective \cite{YellowBook} for frequencies below about $10~\rm mHz$. In comparison, the noise level in the modified dual arm locking configuration has been decreased below the recommended arm locking objective for the entire LISA band, which is well below the TDI capability. Note that due to the $f$ slope in the transfer function of $F_{\rm D}(s)$, the noise limitations at low frequencies in Figure \ref{fig:MDAL_noise} are not maximally suppressed. In LISA a better noise suppression performance could be expected if $F_{\rm D}(s)$ has a high-pass slope steeper than $f$. If the magnitude of the $F_{\rm D}(s)$ filter rolls off faster at low frequencies, the noise floors will become less dependent on the arm length mismatch and asymptotically approach the floor determined by $1/\bar{\tau}$.

\section{Experimental Setup of Realistic Arm Locking Tests}

For the second part of this review, we discuss the experimental verifications of the four arm locking configurations discussed in Section II. In this section we will first discuss the individual components required in our arm locking bench-top experiments, including phasemeters, delay lines, phase-locked loops and arm locking sensors, etc. Then with these components we are able to stabilize the frequency of a numerically controlled oscillator (NCO) to a noisy voltage controlled oscillator (VCO) via any arm locking configuration. Such an arm-locking-based control system is sufficient to demonstrate the noise suppression and closed-loop stability of our arm locking configurations. As a second step, the control system can be integrated with cavity pre-stabilized lasers to provide more noise suppressions. The incorporation of pre-stabilization into arm locking requires an additional actuator to tune the pre-stabilization reference and one of the methods is to use an offset phase-locked loop. The concept of PLL-based arm locking is to stabilize an auxiliary laser, which is phase-locked to a cavity pre-stabilized laser and therefore reproduces the frequency noise property of the master laser. Since the local oscillator of the PLL can be tuned via a frequency or phase modulation, the auxiliary laser can be further stabilized by the arm locking open-loop gain \cite{Wand2009}.

\subsection{Components}

\subsubsection{Phasemeter}
The phasemeter is the most fundamental instrument in LISA's measurement subsystem. The phasemeter precisely measures the heterodyne phase of the laser beat signal. For the phase detection in digital controls, a fast phasemeter with a high bandwidth ($\sim 100~\rm kHz$) is needed to perform phase locking and arm locking. The basic design of the LISA science phasemeter is a digital phase-locked loop (DPLL), which is required to be able to detect the arm length change with a precision of $\sim 1~\rm pm/\sqrt{Hz}$. Therefore, the phase measurement with the $1064~\rm nm$ wavelength is required to have an accuracy of $\sim 1 \mu\rm cycles/\sqrt{Hz}$.

\begin{figure}[tb]
	\centering
		\includegraphics[width=0.49\textwidth]{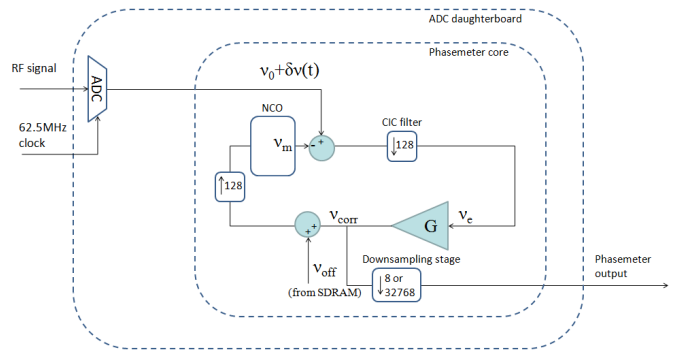}
	\caption{The implementation of the phasemeter on the FPGA. The phasemeter is essentially a ``frequency meter'' that tracks the frequency fluctuations of the input signal.}
	\label{fig:PM_design}
\end{figure}

The phasemeter developed at the University of Florida adapts an architecture of DPLL similar to the LISA phasemeter. As shown in Figure \ref{fig:PM_design}, the phasemeter is implemented on a high-speed FPGA-based digital signal processing (DSP) board. The clock frequency to operate the 14-bit ADC is $f_{\rm clock} = 62.5~\rm MHz$. The digitized heterodyne signal, which carries a frequency $\nu_0 + \delta \nu(t)$ is demodulated by a 48-bit NCO that is phase-locked to the measured signal. The NCO frequency $\nu_{\rm m}$ tracks the frequency of the measured signal, and it is given by the sum of a 16-bit preset offset frequency $\nu_{\rm off}$ and a 48-bit time-varying frequency correction $\nu_{\rm corr}$. The frequency difference between the NCO and the measured signal gives the PLL error $\nu_{\rm e}$. Thus the in-loop frequency variation is given by
\begin{equation}
\nu_{\rm corr} = \frac{G}{1 + G}[\delta\nu(t) + (\nu_0 - \nu_{\rm off})] \approx \delta\nu(t) + (\nu_0 - \nu_{\rm off}).
\end{equation}
Therefore, the phasemeter is essentially a ``frequency meter'', where the frequency fluctuations of the measured signal is faithfully reproduced by $\nu_{\rm corr}$ if $\nu_0 = \nu_{\rm off}$ within the bandwidth. If the preset offset frequency is not exactly equal to $\nu_0$, an offset will be added into $\nu_{\rm corr}$. A typical example of this kind of mismatch is the Doppler frequency error, which is generated from the time-dependent Doppler shift to the returning laser beam.

The noise limitations of this digital phasemeter primarily include a $1/\sqrt{f}$ ADC noise floor and a flat digitization noise floor. Empirically, the ADC noise is given by
\begin{equation}
\delta \varphi_{\rm ADC} (f) = \frac{3\times 10^{-7}}{\sqrt{f/[\rm Hz]}} \cdot \frac{\nu_0}{4~\rm MHz}~\rm cycle~Hz^{-1/2},
\end{equation} 
where $\nu_0$ is the nominal frequency of the input signal.

As previously mentioned by Eq. \ref{Eq:Ndig}, the finite precision of the fixed-point integers of the frequency fluctuations in the phasemeter causes a flat digitization noise in frequency. This is because all the in-loop signals, including the digitized heterodyne signal $\nu_0 + \delta \nu(t)$ and the NCO tracking signal $\nu_{\rm m}$, are registered as fixed-point integers that represent frequencies scaled by the clock frequency $f_{\rm clock}$. The digitization noise in the phasemeter readout due to the round-up and truncation is then given by
\begin{equation}\label{Eq:DIGI_NOISE}
\delta \nu_{dig} (f) = \frac{f_{\rm clock} \cdot 2^{-N}}{\sqrt{6\cdot f_{\rm s}}},
\end{equation}
which is a limited-bandwidth white noise in terms of frequency fluctuations. With the sampling frequency of $62.5~\rm MHz/128 \approx 488~\rm kHz$, the digitization noise of the 48-bit phasemeter is about $1.30\times 10^{-10}~\rm Hz~Hz^{-1/2}$. In some experiments we used a 32-bit phasemeter which increases this to $8.50\times 10^{-6}~\rm Hz~Hz^{-1/2}$.

\subsubsection{Electronic Phase Delay}
A challenging issue in benchtop experiments of LISA interferometry is the simulation of the round-trip propagation between spacecraft. The difficulty of reproducing a LISA-like $33~\rm s$ delay line compromises the validity of LISA interferometry experiments of TDI or arm locking. Compared with unrealistic short delay lines via very long cables or fibers used in most laboratories, one distinctive feature is the emulation of realistic LISA-like delay times and MHz-range Doppler shifts via electronic delay. Such an electronic delay system built with high-bandwidth phasemeters is called an Electronic Phase Delay (EPD) unit \cite{Thorpe2005a}. 

The EPD system is also implemented on an FPGA-based DSP board clocked at $62.5~\rm MHz$. An EPD unit consists of three main components, which are implemented on three parts of the DSP board respectively. The A/D daughtercard is programmed with a 48-bit phasemeter. The phasemeter measures the frequency fluctuation of the digitized laser beat signal with a data rate of $61~\rm kHz$ and then send the data stream to the memory of the motherboard. The motherboard stores the frequency information in a memory buffer for a certain amount of time. The high data rate in the phasemeter ensures the frequency information within the arm locking bandwidth ($\sim \rm kHz$) can be properly delayed by the EPD system. After the delay in the memory buffer, the frequency information is sent to the D/A daughtercard, where an NCO integrates the sum of the frequency fluctuation and the frequency offset to regenerate the delayed copy of the input laser phase. After the NCO output, the digitized signal is converted back to an analog signal by the $500~\rm MHz$ sampling frequency D/A converter. During this routine, a MHz-range Doppler frequency can be added dynamically to the nominal frequency of the delayed signal on the motherboard \cite{Mitryk2012}.

\subsubsection{Arm locking controller}
The arm locking controller is required to provide sufficient gain to suppress the laser frequency noise to below the TDI capability curve within the LISA band. On the other hand, to adequately control the Doppler frequency pulling, the controller is generally required to be AC-coupled with high-pass filtering at very low frequencies ($\sim 10~\mu \rm Hz$) \cite{Thorpe2011}. This design requires at least a few days to measure down to such low frequencies, which is typically difficult for bench-top experiments to achieve. As suggested by Gath \cite{whitepaper}, it is still possible to mitigate the frequency pulling in the presence of a DC-coupled controller if we inject an additional control loop. Such a low-bandwidth loop estimates the Doppler frequency error by measuring the long arm interferometry signals. This information is then feedforwarded to the temperature actuator of the laser to compensate the Doppler frequency error. In UFLIS we have simplified the controller into a purely DC-coupled one without such a feedforward loop, since the Doppler frequency error in our experiments is either negligible or constantly small ($< 10~\rm Hz$).

From the point of view of stability, the design at high frequencies ($> 1~\rm Hz$) is determined by the sensor transfer function since the slope of the controller filter will need to preserve some extra phase at frequencies with a $\sim 90^{\circ}$ phase shift. The expected largest arm length mismatch of LISA is around $\sim 0.5~\rm s$, corresponding to the first null frequency at $\sim 2~\rm Hz$. Therefore, we design an infinite impulse response (IIR) filter with a $1/\sqrt{f}$ slope starting from $\sim 1~\rm Hz$, which is sufficient to maintain enough phase margin even for the longest differential arm length. Such a slope of an IIR filter in the s-domain can be approximately achieved by placing zeros and poles alternately with a frequency spacing ratio of 10, i.e., poles at $1~\rm Hz$, $10~\rm Hz$, $100~\rm Hz$, $1~\rm kHz$ and zeros at $3~\rm Hz$, $30~\rm Hz$, $300~\rm Hz$, $3~\rm kHz$ \cite{Sheard2003}. To convert the coefficients into the z-domain, we perform the bilinear transform with a sampling frequency of $488~\rm kHz$, which is inherited from the data rate of the phasemeter output. Although this filter is designed to be compatible with dual/modified dual arm locking, the demonstration of single/common arm locking experiments may also adapt this filter in the presence of a relatively shorter delay time of $\sim 1~\rm s$, which is comparable to the typical sensor nulls in the dual arm configurations.

The arm locking controller is implemented on the DSP board and directly connected to the sensor output. The frequency fluctuation from the controller output is sent to an numerical controlled oscillator. The latency during the real-time signal processing and data transfer in the sensor/controller is the primary reason that limits the arm locking bandwidth. The specific duration of the latency depends on the complexity of the sensor/controller design. For our arm locking system, the typical duration is on the order of $\sim 10~\rm \mu s$, which yields a bandwidth of a few kHz.  

\subsubsection{Phase-locked Loop}
The phase-locked loop in the optical transponder is required to have a high gain and a high bandwidth. For most experiments such a phase-locked loop can be spared since the EPD unit is capable of delaying an electronic signal by the entire round-trip travel time. The phase-locked loop placed at the far-end is useful when we investigate the effect of the transponder noise on the arm locking performance. Another situation to use the phase-locked loop is the integration of arm locking and cavity pre-stabilization, where a phase-locked loop can be used as a frequency actuator to obtain the tunable reference. In either case, the phase-locked loop is required to have a bandwidth substantially larger than the arm locking loop. We implement an analog phase-locked loop that can adjust the frequency of the Nd:YAG laser with a bandwidth of $\sim 20~\rm kHz$ and the differential frequency noise of the phase-locked loop is considerably lower than that of cavity pre-stabilized lasers.

\subsection{Experimental Setup}

\begin{figure}[tb]
	\centering
		\includegraphics[width=0.49\textwidth]{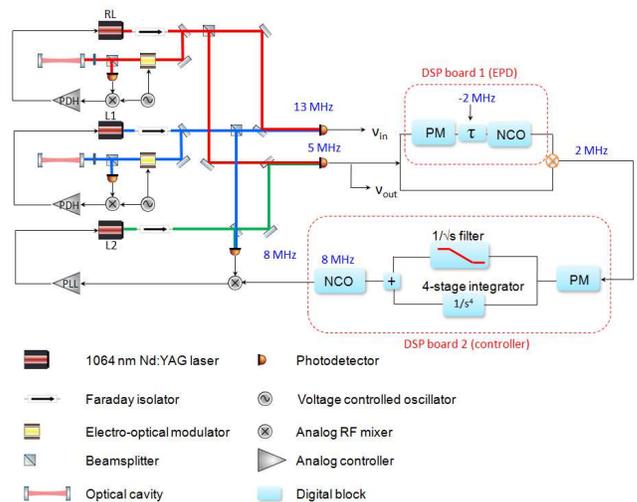}
	\caption{The experimental setup of the single arm locking experiment using an additional heterodyne phase-locked laser. In this setup the reference laser $RL$ and the master laser $L_1$ are cavity stabilized via the Pound-Drever-Hall technique. The slave laser $L_2$ is phase-locked to $L_1$ with a frequency offset, which is driven by the NCO in the arm locking controller. Therefore, within the PLL bandwidth $L_2$ faithfully reproduces the laser frequency noise of $L_1$, both referenced to the ``optical clock'' $RL$. Note that the PLL bandwidth is about $20~\rm kHz$, which is well larger than the arm locking bandwidth ($\sim 1~\rm kHz$); therefore it will not limit our arm locking performance and a direct feedback of the arm locking control signal to the laser is not necessary.}
	\label{fig:SAL_PLL_setup}
\end{figure}

The concept of laser-based arm locking using an offset PLL is to stabilize an auxiliary laser which is heterodyne phase-locked to a cavity pre-stabilized laser such that it obtains both the noise property of the pre-stabilized laser and frequency tunability. In this setup we still assume an ideal optical transponder in each arm and each interferometer output therefore yields the heterodyne beat between the local laser and its identical copy simply delayed by the round-trip time $\tau_i$. The experimental setup of single arm locking is illustrated in Figure \ref{fig:SAL_PLL_setup}, where the master laser $L_1$ is cavity pre-stabilized using the Pound-Drever-Hall technique and the slave laser $L_2$ is phase-locked to $L_1$ with a frequency offset from an NCO. Another cavity pre-stabilized laser $RL$ is used as an optical clock to generate two MHz beat signals with $L_1$ and $L_2$. The $RL-L_2$ beat signal is electronically split into two channels to generate the instantaneous signal and a delayed and Doppler-shifted replica via the EPD unit. The output of LISA's heterodyne interferometer is again simulated by analog mixing of the instantaneous and delayed signals before it is measured by a fast phasemeter. Note that in this setup the phasemeter demodulates the interferometry signal with the exact Doppler shift frequency. The measured frequency fluctuations yield the digitized sensor signal of this single arm locking loop without any Doppler error. The feedback controller then filters the sensor signal and the controller output adjusts the NCO frequency to drive the PLL. 

The function of the PLL is to reproduce the frequency noise $\nu_1$ of the master laser $L_1$ on the slave laser $L_2$, i.e.,
\begin{equation}\label{Eq:HVRAL_PLL}
\nu_2 = \frac{1}{1 + G_0}\nu_2^0 + \frac{G_0}{1 + G_0}(\nu_1 + \nu_{\rm NCO}).
\end{equation}

The first term indicates that the frequency noise $\nu_2^0$ of the free-running $L_2$ is suppressed by the open-loop gain $G_0$. The second term represents the reference noise that this PLL tracks, including the frequency noise $\nu_1$ of the master laser $L_1$ and an additional frequency modulation $\nu_{\rm NCO}$ from the NCO signal. The frequency noise of the NCO is determined by the arm locking loop:
\begin{align}
\nu_{\rm NCO} & = \boldsymbol{S_k} \cdot \begin{bmatrix}\nu_{12}(s)\\\nu_{13}(s)\end{bmatrix} \cdot G_1 \notag\\
							& = (\nu_0 - \nu_2)HG_1,
\end{align}
where $\nu_{1i} = (\nu_0 - \nu_2)(1 - e^{-s\tau_i})$ is the measured frequency noise on each arm with zero transponder noise, $\nu_0$ is the frequency noise of the reference laser $RL$ and $G_1$ is the gain of the arm locking controller.

Substitute the above equation into Eq. \ref{Eq:HVRAL_PLL} and $\nu_2$ becomes
\begin{equation}\label{Eq:HVRAL_NU2}
\nu_2 = \frac{1}{1 + G_0}\nu_2^0 + \frac{G_0}{1 + G_0}[\nu_1 + (\nu_0 - \nu_2)HG_1].
\end{equation}

We combine the terms involving $\nu_2$ to the left and add terms $\nu_0 - \frac{1}{1 + G_0}\nu_0 - \frac{G_0}{1 + G_0}\nu_0 (= 0)$ to the right. Then Eq. \ref{Eq:HVRAL_NU2} can be simplified into
\begin{align}\label{Eq:HVRAL_TF}
& \left( 1 + \frac{G_0}{1 + G_0}HG_1 \right)(\nu_2 - \nu_0) \notag\\
& = \frac{1}{1 + G_0}(\nu_2^0 - \nu_0) + \frac{G_0}{1 + G_0}(\nu_1 - \nu_0).
\end{align}

If we bring the factor of $( 1 + \frac{G_0}{1 + G_0}HG_1 )$ on the left side to the right, the first term on the right indicates that the frequency noise of a free-running $L_2$ relative to the reference laser is double suppressed by both the open-loop gain of the PLL and that of the arm locking loop. The second term representing the reproduction of the frequency noise of $L_1$ relative to the reference laser, is now suppressed by the open-loop gain of arm locking $\frac{G_0}{1 + G_0}HG_1$.

From this equation we obtain the equivalent open-loop transfer function of the entire system
\begin{equation}
TF_{\rm OL} = \frac{G_0}{1 + G_0}HG_1.
\end{equation}

Since the bandwidth of the PLL ($\sim 20~\rm kHz$) is well above the bandwidth of the arm locking loop, we can always assume $1/(1+G_0) \approx 0$ and $G_0/(1 + G_0) \approx 1$. Then Eq. \ref{Eq:HVRAL_TF} can be further reduced into
\begin{equation}   
TF_{\rm CL} = \frac{1}{1 + HG_1} = \frac{\nu_2 - \nu_0}{\nu_1 - \nu_0}.
\end{equation}

The expression of the closed-loop transfer function indicates that the input frequency noise of the laser beat signal $RL-L_2$ is directly suppressed by the arm locking open-loop gain. In comparison with the LISA situation, the frequency noise $\nu_1 - \nu_0$ represents that of the pre-stabilized laser, i.e., $\nu_{L_1}^0(s)$ in Eq. \ref{Eq:RALM_LFN}. The output frequency noise $\nu_2 - \nu_0$ represents that of the arm locked laser $\nu_{L_1}(s)$. In both cases, the noise suppression is given by the closed-loop gain $1/(1 + HG_1)$.

\section{Experimental Results of Realistic Arm Locking}

\subsection{Single/Common Arm Locking}

\subsubsection{Single arm locking with cavity pre-stabilized lasers}

\begin{figure}[tb]
	\centering
		\includegraphics[width=0.45\textwidth]{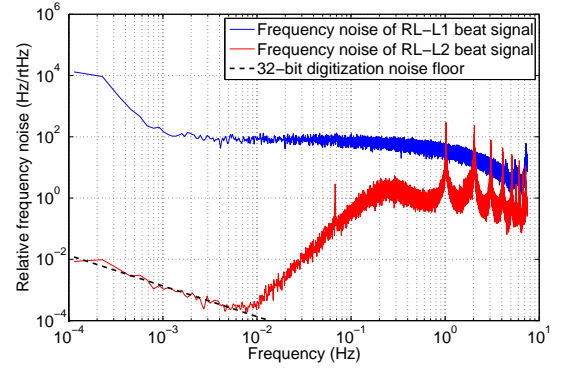}
	\caption{The noise spectra of the initial $RL-L_1$ beat signal (blue) and the stabilized $RL-L_2$ beat signal (red). Compared with the frequency noise spectrum of initial $RL-L_1$ beat signal, the frequency noise spectrum of stabilized $RL-L_2$ beat signal is suppressed by the 4-stage integrators by 5 to 6 orders of magnitude from $0.1~\rm mHz$ to $10~\rm mHz$.}
	\label{fig:SAL_PLL_spectrum}
\end{figure}

We measured the frequency noises of the $RL-L_1$ beat signal and the $RL-L_2$ beat signal in Figure \ref{fig:SAL_PLL_setup} as the input noise and the output noise, respectively. Their noise spectra in the low frequency region are shown in Figure \ref{fig:SAL_PLL_spectrum}. In the low frequency region the noise suppression is limited by the same $1/f$ slope due to the 32-bit digitization noise $\delta \nu_{\rm dig}$ in the arm locking controller. The white digitization noise determined by \ref{Eq:Ndig} is equivalent to an additional term in the sensor signal. The closed-loop transfer function from the error point to the output is given by $G/(1 + GH_{\rm S}) \approx 1/H_{\rm S} \approx 1/\tau s$. Therefore, the digitization noise is accumulated by the closed-loop and the noise floor scales with $1/\tau$:
\begin{equation}\label{Eq:SAL_DNF}
N_{\rm{dig}}(f) = \frac{\delta \nu_{\rm dig}}{\tau (2\pi f)} = \frac{1.35 \times 10^{-6}}{f}~\rm{Hz~Hz^{-1/2}},
\end{equation}
where the 32-bit digitization noise $\delta \nu_{\rm dig}$ is calculated using Eq. \ref{Eq:DIGI_NOISE}. Eq. \ref{Eq:SAL_DNF} indicates that similar to the transponder noise and Doppler errors, the digitization noise in the arm locking controller is also integrated by the arm locking loop and scaled by the light travel time. Note that the PLL in this setup also introduces spurious phase variations $\delta\varphi_{\rm PLL}$, which is well suppressed by the arm locking open-loop gain together with the reference signal of the PLL.

In addition to the demonstration of auxiliary phase-locked laser as the tunable reference, the single arm locking setup has also been integrated with cavity pre-stabilized lasers using other tunable references. One approach is to replace the fixed optical cavity with a piezoelectric transducer (PZT) actuated cavity \cite{Mohle2013}, where the central frequency of the stabilized laser can be tuned by adjusting the length of the cavity. Another approach uses a broadband electro-optical modulator (EOM) and the RF sideband is locked to the fixed-length cavity \cite{Livas2009}. The local oscillator driving the EOM is tuned by the arm locking feedback signal; therefore, the tuning of the modulation/demodulation RF signal allows the tuning of the central frequency of the pre-stabilized laser.

\subsubsection{Common arm locking}

\begin{figure}[tb]
	\centering
		\includegraphics[width=0.49\textwidth]{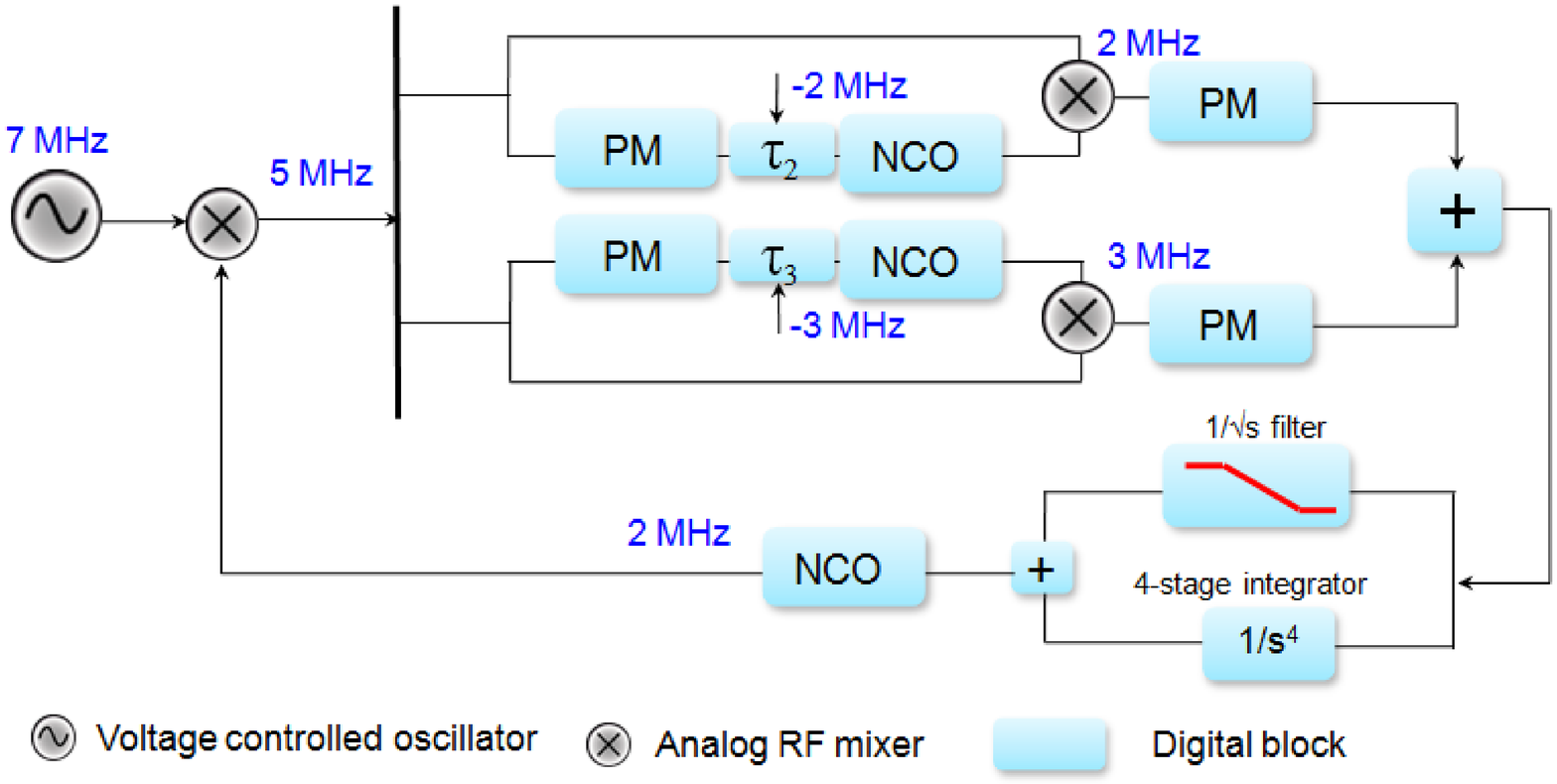}
	\caption{The experimental setup of common arm locking using an NCO to track the input noise. The delay times are $2.1~\rm s$ and $1.9~\rm s$.}
	\label{fig:CAL_NCO_setup}
\end{figure}

Although common arm locking takes a linear combination of the phasemeter measurements on both arms, it still resembles single arm locking in a variety of ways, such as the irrelevance to the arm length mismatch. Here we first demonstrate the validity of common arm locking with relatively short time delays ($\sim 2~\rm s$). As shown in Figure \ref{fig:CAL_NCO_setup}, the common arm sensor is implemented on two DSP boards which function respectively as front end and back end. The front end includes two EPD units that simulate the round-trip travel time on two different arms individually. We use a VCO as a noisy oscillator for initial experiments. The VCO signal is electronically split into two arms and the signal on each arm is split again to generate a prompt signal and a delayed and Doppler shifted signal via the EPD unit. On the other DSP board the back end starts with two phasemeters that measures the phase difference on each arm individually. The mapping vector calculates the sum of the two phase measurements and then send the 32-bit error signal into the controller filter. In this experiment the delay times are set to be $2.1~\rm s$ and $1.9~\rm s$. The corresponding sensor has nulls starting from $1/(\Delta \tau) = 10~\rm Hz$, which is already beyond the LISA band; however, there are also local minima at multiples of $1/\bar{\tau} = 0.5~\rm Hz$ where the phase shift is still close to $\pi/2$ and unsuppressed noise peaks should still be expected.

\begin{figure}[tb]
	\centering
		\includegraphics[width=0.45\textwidth]{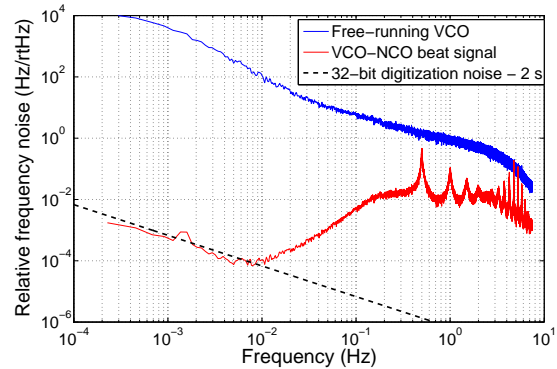}
	\caption{The noise spectra of the free-running VCO signal (blue) and the VCO-NCO beat signal (red). The noise spectrum of the residual frequency noise is limited by the 32-bit digitization noise floor, which is given by the integrated quadrature sum of the 32-bit digitization noise from the two independent phasemeter.}
	\label{fig:CAL_NCO_spectrum}
\end{figure}

The measurement results of the original noise and residual noise is shown in Figure \ref{fig:CAL_NCO_spectrum}. Due to the local minima at $n/\bar{\tau}$ in the sensor magnitude response, the residual frequency spectrum still exhibits periodic peaks, although not noise enhancements, at these frequencies. In the low frequency region, the noise suppression is again limited by a $1/f$ slope due to the 32-bit digitization noise sent to the controller. The controller receives the quadrature sum of the 32-bit digitization noise from both phasemeters. Therefore the digitization noise floor is given by
\begin{equation}
N_{\rm dig}(f) = \frac{\sqrt{2}\delta \nu_{\rm dig}}{\bar{\tau}(2\pi f)} = \frac{9.53 \times 10^{-7}}{f}~\rm{Hz~Hz^{-1/2}}.
\end{equation}
Compared with the single arm locking experiments, the digitization noise floor has been decreased by a factor of $\sqrt{2}$ due to the longer average delay time of $2~\rm s$.

\subsection{Dual/Modified Dual Arm Locking}

\begin{figure*}[tb]
	\centering
		\includegraphics[width=0.95\textwidth]{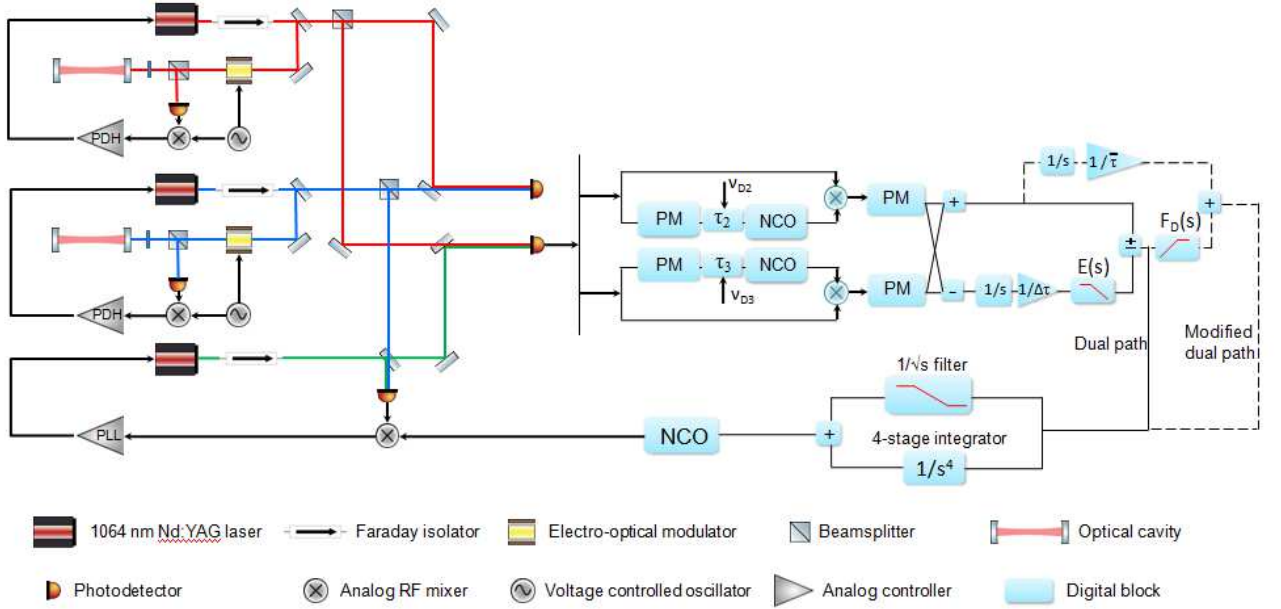}
	\caption{Experimental setup of dual arm locking (the solid path) and modified dual arm locking (the dashed path) with cavity pre-stabilized lasers using an auxiliary phase-locked laser. We measured the frequency noise of the arm locking stabilized $RL-L_2$ beat signal and cavity pre-stabilized $RL-L_1$ beat signal.}
	\label{fig:ALSetup}
\end{figure*}

\subsubsection{Dual arm locking}

Figure \ref{fig:ALSetup} illustrates the experimental setup of dual/modified dual arm locking integrated with cavity pre-stabilized lasers. Compared to common arm locking, the implementation of the dual arm locking sensor requires an additional differential path of the two phase measurements. In the differential path, the differential frequency fluctuations are integrated and scaled by $1/\Delta \tau$. A low-pass filter with a single pole at $\sim 1~\rm Hz$ is placed in the differential path to attenuate the excessive phase loss. The common and differential paths are subsequently added to generate the dual arm locking sensor signal. To investigate the influence of varying arm lengths on the noise performance, we measured the frequency noise spectra of the arm locking stabilized laser ($RL-L_2$ beat signal) under the circumstances of a long arm length mismatch ($\sim 1.5\%$) and a short arm length mismatch ($\sim 0.15\%$). The other parameters configured in the experiments are listed in Tab \ref{tab:DualParams}.

\begin{table}
\caption{\label{tab:DualParams}Parameters in dual arm locking experiments}
\begin{center}
\begin{tabular}{lccc}
\hline
\hline
Parameter&Symbol&Value&Units\\
\hline
Sampling frequency&$f_{\rm s}$&488&$\rm kHz$\\
Averaged delay time&$\bar{\tau}$&$33$&$\rm s$\\
Differential delay time&$\Delta \tau$&$0.25,\, 0.025$&$\rm s$\\
Doppler shift on arm 1-2&$\nu_{\rm D2}$&$-2$&$\rm MHz$\\
Doppler shift on arm 1-3&$\nu_{\rm D3}$&$-3$&$\rm MHz$\\
Controller gain slope&$G(s)$&$f^{-0.5}$&$ $\\
Unity gain frequency&$f_{\rm UGF}$&$1$&$\rm kHz$\\
Phase margin&$\Delta \phi$&$\approx 45$&$\rm degree$\\
\hline
\hline
\end{tabular}
\end{center}
\end{table}

In the presence of a short differential delay time like $0.025~\rm s$, a relatively high noise floor ($\sim 10^{-5}/f~\rm Hz~Hz^{-1/2}$) caused by the 32-bit digitization noise in phasemeters would be expected. For a differential delay time less than $1~\rm ms$ ($\Delta L \sim 300~\rm km$), the residual noise would fail to meet the $0.3~\rm Hz~Hz^{-1/2}$ objective. This potential issue indicates that a controller with 32-bit fixed-point precision is not always ideal for the dual arm locking configuration. For this reason, we enhance the precision of the sensor/controller to 48-bit to decrease the digitization noise by a factor of $2^{16}$.

\begin{figure}
	\centering
		\includegraphics[width=0.45\textwidth]{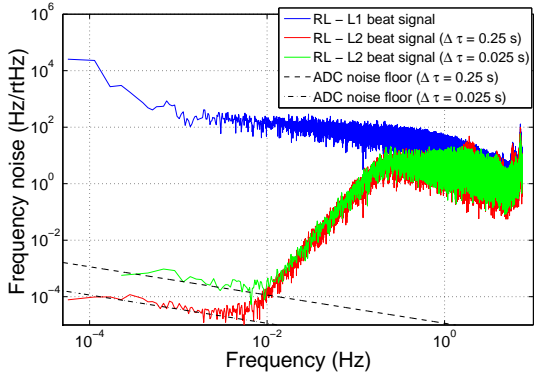}
	\caption{The noise spectra of the cavity stabilized beat signal $RL-L_1$ (blue) and the beat signal $RL-L_2$ further stabilized by dual arm locking. Compared to the $RL-L_1$ beat signal, the frequency noise of $RL-L_2$ is suppressed by 7 to 8 orders of magnitudes in the noise limited region, where the dominant noise floor comes from the phasemeter ADCs. Also we investigate the frequency noise spectra of $RL-L_2$ in the presence of various differential delay times. These measurements simulate the inversely proportional change of the limiting noise floor in accordance with the change of the arm length mismatch in dual arm locking.}
	\label{fig:DAL_PLL_spectrum}
\end{figure}

Figure \ref{fig:DAL_PLL_spectrum} shows the measured frequency noise of the cavity stabilized laser ($RL-L_1$ beat signal) and the arm locking stabilized laser ($RL-L_2$ beat signal) with different differential delay times. The linear spectral densities of the frequency noise are calculated in the steady state, in order to remove the effect of initial transients (with a duration of approximately $400~\rm s$). The red curve represents the frequency noise of the $RL-L_2$ beat signal in the case of $0.25~\rm s$ differential delay time. The first noise enhancement peak is located at around $1/(2\Delta \tau) = 2~\rm Hz$ as expected. Compared to the $RL-L_1$ beat signal, the frequency noise of the $RL-L_2$ beat signal has been suppressed by 7 to 8 orders of magnitude at low frequencies. The green curve represents the frequency noise of the $RL-L_2$ beat signal in the case of $0.025~\rm s$ differential delay time, which corresponds to a higher noise level.

The performances of both two cases are gain limited at frequencies above about $10~\rm mHz$. At lower frequencies, the noise spectra of the two cases are limited by different $1/f^{1/2}$ noise slopes, as indicated by the dashed lines. These $1/f^{1/2}$ noise floors are caused by the ADC noise in the phasemeter on each arm coupled into the dual arm locking loop. The phasemeter ADC noise, which is primarily attributed to the temperature-dependent phase dispersion, has a $1/f$ slope in the power spectral density as a phase noise. Therefore, when represented in the LSD of a frequency noise, the phasemeter ADC noise has a $f^{1/2}$ slope. Empirically, this frequency noise is measured to be
\begin{equation}\label{Eq:RALM_ADC}
\delta \nu_{\rm ADC}(f) =  7.5\times 10^{-8} \cdot \frac{2\pi\nu_0}{1~\rm MHz} \sqrt{f}~{\rm Hz~Hz^{-1/2}},
\end{equation}
where $\nu_0$ is the nominal RF frequency of the digitized signal.

As detailed in Eq. \ref{Eq:RALM_LFN}, the ADC noises in the two phasemeter channels will be added in quadrature and multiplied with $\frac{h_+ + h_-}{H}$. The integrator in the differential channel makes it dominate in the low frequency range. Therefore, the ADC noise floor is given by
\begin{equation}
N_{\rm ADC}(f) = \frac{\sqrt{\delta \nu_{\rm ADC}^2|_{3~\rm MHz} + \delta \nu_{\rm ADC}^2|_{2~\rm MHz}}}{2(2\pi f)\Delta \tau}.
\end{equation}

For the case of $\Delta \tau = 0.25~\rm s$, the expected noise floor is about $1.15\times 10^{-5} f^{-1/2}~\rm Hz~Hz^{-1/2}$ and for $\Delta \tau = 0.025~\rm s$ the noise floor is higher by a factor of 10. The observed noise floors in Figure \ref{fig:DAL_PLL_spectrum} agree very well with the theoretical values. This result therefore demonstrates that the noise limitation in the dual arm locking configuration is inversely proportional to the differential arm length, as theoretically calculated by McKenzie \etal \cite{McKenzie2009}.   

\subsubsection{Modified dual arm locking}

\begin{figure}
	\centering
		\includegraphics[width=0.45\textwidth]{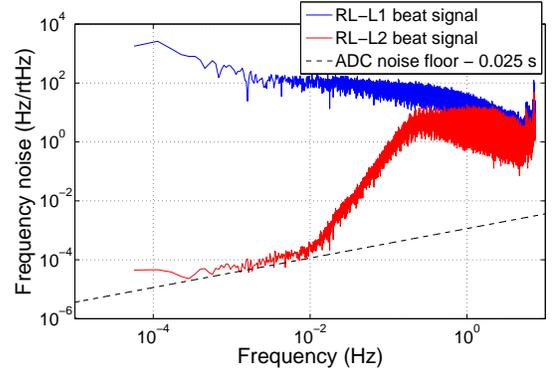}
	\caption{Measured frequency noise of the cavity stabilized laser and the modified dual arm locking stabilized laser in the presence of a $0.15\%$ arm length mismatch. Compared with the previous dual arm locking experiment, the noise suppression performance has been siginificantly improved. The $1/f^{1/2}$ slope ADC noise floor in Figure \ref{fig:DAL_PLL_spectrum} has been surpassed by a $f^{1/2}$ slope, which is determined by the transfer function of the modified dual arm locking sensor as well as the arm length mismatch. The noise floor is given by Eq. \ref{Eq:MDAL_Noise}.}
	\label{fig:MDAL_PLL_spectrum}
\end{figure}

The specific design of the modified dual arm locking sensor is already described in Section II.C.4. As shown by the dashed path in Figure \ref{fig:ALSetup}, in the low frequency region the dual component is suppressed by a high-pass filter $F_{\rm D}(s)$ with a zero at DC and a pole at around $30~\rm mHz$. The common component is integrated and scaled by $1/\bar{\tau}$. Figure \ref{fig:MDAL_PLL_spectrum} shows the measurement under the circumstance of a short differential arm length of $0.15\%$ ($0.025~\rm s$). The other parameters are configured as listed in Tab \ref{tab:DualParams}. Compared with the dual arm locking result with the same differential arm length, the frequency noise of the arm locking stabilized signal has been further mitigated down to a different noise floor with a $f^{1/2}$ slope. The ADC noise given by Eq. \ref{Eq:RALM_ADC} is still the dominant noise source in this modified dual arm locking loop. In the noise limited region ($0.1~\rm mHz$ - $10~\rm mHz$), the differential channel still dominates in the presence of a short differential arm length like $0.15\%$. However due to the high-pass filter $F_{\rm D}(s)$, the transfer function of the differential channel becomes a constant factor of $\bar{\tau}/(2\pi\Delta\tau)$. Consequently, based on Eq. \ref{Eq:RALM_LFN} the ADC noise floor is given by
\begin{equation}\label{Eq:MDAL_Noise}
N_{\rm ADC}(f) = \sqrt{\delta \nu_{\rm ADC}^2|_{3~\rm MHz} + \delta \nu_{\rm ADC}^2|_{2~\rm MHz}} \cdot \frac{\bar{\tau}}{2\pi\Delta \tau H}.
\end{equation}
For the case of $\Delta \tau = 0.025~\rm s$, the expected noise floor is about $5.73\times 10^{-4} f^{1/2}~\rm Hz~Hz^{-1/2}$ and the observed noise floor matches closely to the theoretical value.   

\begin{figure*}[t]
	\centering
		\includegraphics[width=0.95\textwidth]{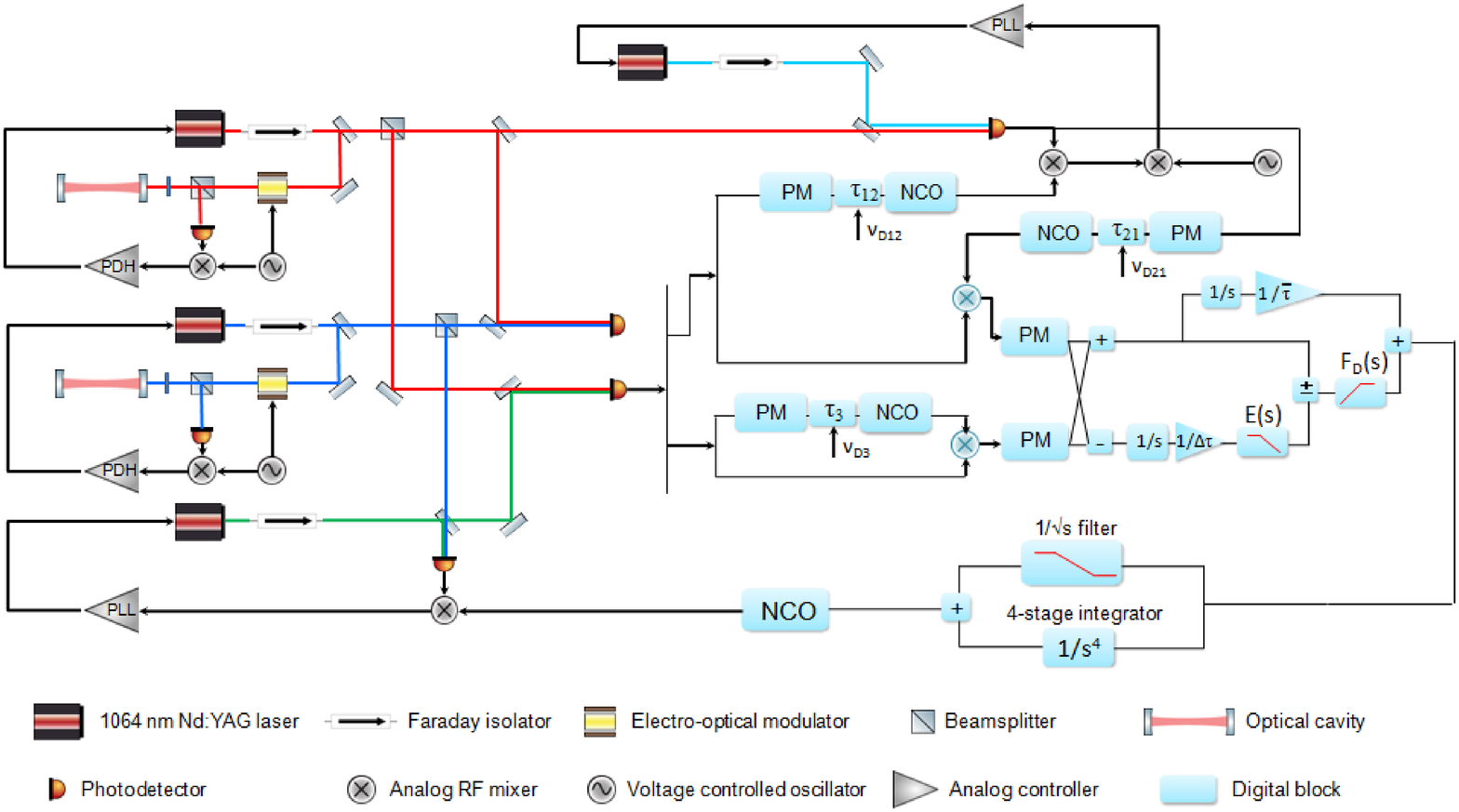}
	\caption{Experimental setup of modified dual arm locking with an optical transponder. The delay line to simulate the LISA arm 1-2 is divided equally into two, representing the outgoing and return travel time individually. Another beat signal $RL - L_3$ is phase-locked to the $RL-L_2$ delayed by the outgoing travel time with an offset frequency of $2~\rm MHz$ and then delayed by the return travel time to have the heterodyne interference with the prompt beam. In this setup the voltage controlled oscillator demodulating the heterodyne frequency of the PLL has an equivalent function as the clock on far spacecraft, where the clock noise enters the phase measurement of the OPLL in a similar way.}
	\label{fig:ALSetup_OPLL}
\end{figure*}

\subsection{Integration with Optical Transponders}

With the results achieved in the previous experiments, we take one step forward to the aim of realistic arm locking hardware simulation by introducing the optical transponder noise at far spacecraft. We implement a real phase-locked loop at the far-end rather than assume that the optical transponder functions perfectly with an infinite feedback gain. To simulate an additional noise source from the far spacecraft, we need to divide the delay time at the EPD unit equally and delay the input electronic signal using two split delay lines in cascade. Analogously, the arm length change due to the LISA spacecraft motion is insignificant over a short time interval of $33~\rm s$, which makes the outgoing time and the return time are almost the same. Also, since the common and differential noise between two LISA arms are both given by the quadrature sum, we can apply the additional transponder noise on only one arm without loss of quantitative validity; otherwise a 5th laser would be required to simulate the other far spacecraft.

The experimental setup to verify the transponder noise floor in modified dual arm locking is illustrated by Figure \ref{fig:ALSetup_OPLL}. In this setup we still use an auxiliary phase-locked laser $L_2$ to obtain tunability of the pre-stabilization reference. As the outgoing beam from the ``local spacecraft'', the beat signal $RL-L_2$ is electronically split and sent to the corresponding ``far spacecraft''. We split the round-trip time of $33.025~\rm s$ equally into two delay lines, representing the outgoing travel time $\tau_{12}$ and the return travel time $\tau_{21}$. Once the $RL-L_2$ beat signal is delayed by $\tau_{12}$ and arrives at the ``far spacecraft'', its delayed and Doppler-shifted version (a $3~\rm MHz$ NCO signal) is used to heterodyne phase-lock a ``far laser'' represented by the $RL - L_3$ beat signal. The $5~\rm MHz$ $RL - L_3$ beat signal tracks the frequency noise of $RL - L_2$ and also carries uncorrelated noise, which includes the analog electronic noise (from the photodetector, mixer, etc.) as well as the residual noise due to the finite gain of the PLL controller. The $2~\rm MHz$ offset frequency of the PLL, which is driven by a function generator, is synchronized to the master clock to avoid any unwarranted Doppler frequency errors. The frequency noise of the function generator signal enters the PLL and becomes a part of the transponder noise. This process resembles the clock noise that enters the far-end PLL during the demodulation in the phase measurement. Then the phase-locked $RL - L_3$ beat signal is delayed by the return travel time $\tau_{21}$ using another EPD channel and form the long arm interferometry with the ``local beam'' $RL - L_2$ beat signal.

In this setup, primary contributions to the transponder noise includes the limited PLL gain, the analog electronic noise $\nu_{\rm e}$ and the frequency noise $\nu_{\rm CL}$ of the function generator:
\begin{align}
\delta \nu_{\rm Trans} & = \frac{1}{1 + G_2}(\nu_0 - \nu_3^0)e^{-s\tau_{21}} + \frac{G_2}{1 + G_2}\nu_{\rm e}e^{-s\tau_{21}} \notag\\
											 & + \frac{G_2}{1 + G_2}\nu_{\rm CL}e^{-s\tau_{21}},
\end{align}
where $G_2$ is the open-loop gain of the far-end PLL. Following a similar procedure as in the previous section, the transponder noise floor is given by
\begin{align}
N_{\rm Trans} & = \frac{1}{1 + G_0}\frac{(h_+ + h_-)G_1}{1 + \frac{G_0}{1 + G_0}HG_1}\delta \nu_{\rm Trans} \notag\\
									 & \approx \frac{h_+ + h_-}{H}\delta \nu_{\rm Trans} \qquad (G_0, \, G_1 \gg 1).
\end{align}
 
This result is consistent with the second term of the result in Eq. \ref{Eq:RALM_LFN}, proving the validity of this experimental setup to demonstrate the transponder noise floor. With a certain controller gain, the transponder noise is measured to be approximately $3\times 10^{-4}~f^{1/5}~\rm Hz~Hz^{-1/2}$. The dominant noise comes from the phase-locked loop, while the frequency noise of the function generator signal is much lower. Here we have tuned the PLL into the gain limited region such that the transponder noise floor can be proportionally manipulated by adjusting the gain of the PLL controller.

\begin{figure}[t]
	\centering
		\includegraphics[width=0.45\textwidth]{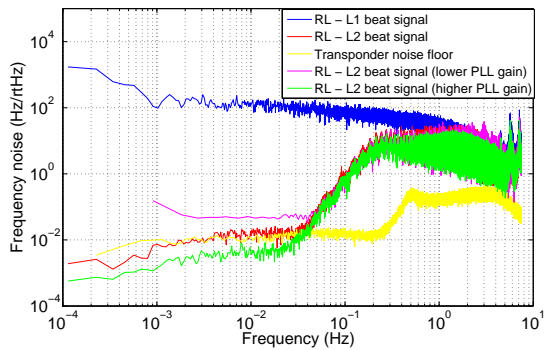}
	\caption{Noise spectra of the laser beat frequency stabilized by modified dual arm locking in the presence of transponder noise. In this figure the stabilized frequency noise is represented by the red curve, which is limited by the transponder noise floor. The yellow curve represent the transponder noise floor given by the combined noise multiplied with the differential arm gain $\bar{\tau}/(2\pi \Delta \tau \cdot H)$.As the PLL noise increases or decreases, the transponder noise floor in the stabilized laser frequency also tracks this change, which is shown by the green (a higher PLL gain) and purple (a lower PLL gain) curves.}
	\label{fig:MAL_PLL_spectrum}
\end{figure}

The measurement results are illustrated in Figure \ref{fig:MAL_PLL_spectrum}. As the differential delay time equals $0.025~\rm s$, the expected transponder noise floor below $30~\rm mHz$ is approximately given by $0.20~f^{1/5}~\rm Hz~Hz^{-1/2}$, which limits the noise suppression performance in that region. Our measurement has originally demonstrated that the arm locking stabilized frequency noise agrees with this expected noise floor and still sufficiently meet the LISA requirement in the presence of transponder noise. For frequencies around $30~\rm mHz$ the arm locking performance is still gain limited and the transponder noise floor given by $1/\Delta \tau s$ is still below the $RL-L_2$ frequency noise. We also have demonstrated that as the PLL gain changes, the corresponding transponder noise will track the change accordingly within the gain limited range. The measurements indicate that the transponder noise floor inversely scales with the PLL gain as expected.

\subsection{Summary}
The measurement results of various arm locking configurations have demonstrated substantial noise suppressions compared to the cavity pre-stabilized laser. The noise suppression performance of arm locking depends on the amplitude of laser frequency noise and transponder noise, geometry of the LISA orbits and consequent sensor/controller design. The measurements have indicated that for LISA the single/common arm locking configuration is independent of the arm length mismatch yet their sensor nulls at low frequencies would severely limit the controller gain and cause noise peaks inside the LISA band. In comparison, dual arm locking allows a more flexible controller design due to the flat transfer function through the entire LISA band. However, the disadvantage of dual arm locking becomes evident in the presence of a short arm length mismatch, which significantly amplifies the transponder noise floor. Modified dual arm locking is a compromise between common and dual arm locking, featuring the advantages of both. The final noise floor of modified dual arm locking is determined by the arm length mismatch, as well as the specific design of the sensing filters.

\section{Doppler Frequency Pulling}

\begin{figure}
	\centering
		\includegraphics[width=0.45\textwidth]{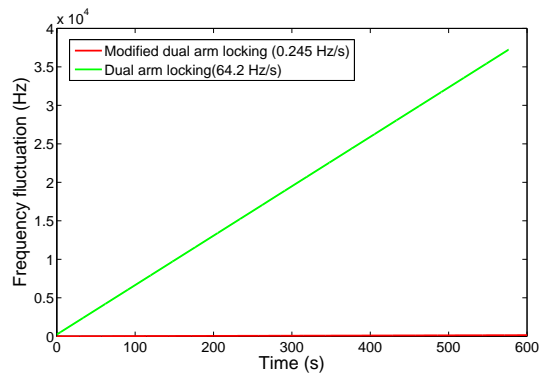}
	\caption{\label{fig:Doppler_DAL_frequency} Observed frequency pulling of dual/modified dual arm locking with Doppler frequency errors. For a dual arm locking and a modified dual arm locking configuration, the observed frequency pulling rate is $64.2~\rm Hz/s$ and $0.245~\rm Hz/s$, respectively. These measurement results are consistent with the theoretical predictions and demonstrate that the modified dual arm locking sensor is capable of alleviating the Doppler issue substantially.}
\end{figure}

In addition, we have verified the frequency pulling of dual/modified dual arm locking in the presence of Doppler frequency errors. Based on the same setup, we introduce constant Doppler frequency errors into the phase measurements by shifting the clock frequency. With a frequency shift of $2~\rm kHz$ in the $62.5~\rm MHz$ clock, the frequency errors generated from $2~\rm MHz$ and $3~\rm MHz$ Doppler shifts are $6.4~\rm Hz$ and $9.6~\rm Hz$, respectively. This constant Doppler frequency error was used to demonstrate the frequency pulling rate under different arm locking configurations. With $\bar{\tau} = 33~\rm s$ and $\Delta\tau = 0.025~\rm s$, the expected frequency pulling rates are 
\begin{align}
\left(\frac{\rm{d}\nu_{\rm{L}}}{\rm{d}t}\right)_{\rm{Dual}} & = \frac{\Delta \nu_{\rm D-}}{2\Delta \tau} = 64~\rm Hz/s, \notag\\
\left(\frac{\rm{d}\nu_{\rm{L}}}{\rm{d}t}\right)_{\rm{Mod\,Dual}} & = \frac{\Delta \nu_{\rm D+}}{2\bar{\tau}} = 0.24~\rm Hz/s.
\end{align}

The observed frequency pulling for both two cases is shown in Figure \ref{fig:Doppler_DAL_frequency}. In a duration of $600~\rm s$, the output frequency has drifted by more than $35~\rm kHz$ when dual arm locking is used. In contrast, the modified dual arm locking configuration limits the 1-hour frequency pulling within a range of less than $900~\rm Hz$ and this frequency pulling rate is even smaller than a typical drift rate of cavity stabilized lasers. From the frequency data we have obtained the frequency pulling rates are $64.2~\rm Hz/s$ and $0.245~\rm Hz/s$. The observed frequency pulling rates match the expected values and demonstrated that the modified dual arm locking sensor is capable of alleviating the Doppler issue substantially.

\section{Conclusion}

In this paper we reviewed the control system as well as the performance of arm locking in LISA. In the experiments we tested advanced arm locking schemes for LISA using our LISA interferometer testbed. The EPD-based electro-optical arm locking hardware simulations can effectively and faithfully reproduce realistic LISA-like conditions such as $33~\rm s$ light travel time and variable MHz Doppler shifts, which are vital for the validity of arm locking experiments. In particular, the dual and modified dual arm locking configurations are linear combinations of LISA inter-spacecraft phase measurements optimized for the noise performance and minimization of the Doppler-induced frequency pulling issue. The incorporation of advanced arm locking schemes with cavity stabilized lasers has demonstrated that arm locking can be easily reconciled with the cavity stabilization without explicitly degrading the noise suppression performance of either of them. In addition, a more realistic demonstration of modified dual arm locking in the presence of optical transponder noise is also presented. The transponder noise limitation observed in the stabilized laser noise can be equivalently considered as a manifestation of any noise source presented in the optical transponder as they all couple into the arm locking control system in the same fashion. The experiment also reveals that in the presence of a non-negligible transponder noise and a relatively short arm length mismatch ($\Delta \tau = 0.025~\rm s$), our modified dual arm locking configuration with cavity stabilization still sufficiently meets the TDI capability with a margin of more than $25,000$ at $3~\rm mHz$. Such a frequency-stabilized laser with extremely low frequency noise could reduce the complexity of the ranging subsystem and even relax the burden on the data processing of TDI. Our result indicates that with the help of arm locking, the requirements on the design of other IMS subsystems can be less stringent to considerable degrees.

\section*{Acknowledgement}

The authors thank especially Josep Sanjuan, Vinzenz Wand and the entire UF LISA group for helpful discussions and experimental support. We also thank the former LISA project team, especially Kirk McKenzie, Daniel Shaddock, James Ira Thorpe and Jeffery Livas for their original and important contributions. This work was supported by NASA Grant NNX09AF99G.

\bibliography{ref}

\end{document}